\documentclass{aa}  

\usepackage{graphicx}
\usepackage{hyperref}
\usepackage{txfonts}
\usepackage{natbib}
\bibpunct{(}{)}{;}{a}{}{,}
\newcommand{\um}{$\mu$m}

\def\arcsec{\hbox{$^{\prime\prime}$}}
\def\utw{\smash{\rlap{\lower5pt\hbox{$\sim$}}}}
\def\udtw{\smash{\rlap{\lower6pt\hbox{$\approx$}}}}

\def\Lsun{\hbox{\it L$_\odot$}}

\def\Msun{\hbox{\it M$_\odot$}}

\def\Teff{\hbox{\it T$_{\rm eff}$}}

\def\Mk{\hbox{\it M$_{\rm K}$}}
\def\Mbol{\hbox{\it M$_{bol}$}}
\newcommand{\Ks}{{\it K$_{\rm s}$}}

\newcommand{\Aks}{{\it A$_{\rm K_{\rm s}}$}}
\newcommand{\Ak}{{\it A$_{\rm K}$}}

\def\BCK{\hbox{\it BC$_{\rm K}$}}
\def\BCJ{\hbox{\it BC$_{\rm J}$}}
\def\BCH{\hbox{\it BC$_{\rm H}$}}
\def\BCKs{\hbox{\it BC$_{\rm K_S}$}}

\def\simgr{\mathrel{\hbox{\rlap{\hbox{\lower4pt\hbox{$\sim$}}}\hbox{$>$}}}}

\def\vlsr{\hbox{$v_{LSR}$}}

\begin{document}

   \title{Candidate red supergiants from Gaia DR3 BPRP spectra:
   From the Perseus to the Scutum-Crux spiral arms  }

   \subtitle{}

\author{Maria Messineo
          \inst{1,2}
           }

   \institute{Dipartimento di Fisica e Astronomia 
“Augusto Righi”, Alma Mater Studiorum, Università di Bologna, Via Gobetti 93/2, 
I-40129 Bologna, Italy\\
              \email{maria.messineo@inaf.it}
         \and
             INAF - Osservatorio di Astrofisica e Scienza dello Spazio 
di Bologna, Via Gobetti 93/3, I-40129 Bologna, Italy\\
             }

   \date{Received September 15, 1996; accepted March 16, 1997}

   \abstract
   {Our position within the Galactic plane and the dust obscuration make 
   it challenging to retrieve 
   a true picture of the Milky Way’s morphology. 
   While the Milky Way has been recognized as a barred spiral galaxy 
   since the 1960s, there is still uncertainty about the exact 
   number of spiral arms it contains. 
   Currently, our understanding of the Galaxy is evolving 
   thanks to the unprecedented detail provided by Gaia’s 
   parallactic distances.}
   {To  shed light on the spatial distribution of red supergiants (RSGs) on the Disk
   and  their uniformity of parameters across it,
   a census of Galactic RSGs detected by Gaia is needed.}
   {Candidate RSGs were extracted 
   from the combined Gaia DR3 and 2MASS catalogs using
   color criteria and parallactic distances. 
   The sample includes 335 stars that were not included in 
   catalogs of previously known RSGs detected by Gaia DR3.
   Interstellar and circumstellar
   extinction values were estimated from the infrared bands.
   Spectral types were collected from Simbad or VIZIER databases
   and, for 135 candidates, were inferred  from the Gaia DR3 
   BPRP spectra. Stellar luminosities were inferred using 
   photometric measurements and the Gaia DR3  distances.}
   {The analysis yielded a genuine sample
   of O-rich late-type stars, and the calculated luminosities 
   confirm that the sample is mostly made of stars brighter than  \Mbol=-5 mag.
   This new sample represents a 40\% increase in 
   the number of highly probable RSGs  compared to  previous studies.
   When looking at the X and Y distribution on the Galactic 
   plane, beside the populous Perseus associations of RSGs
   and the Sagittarius group of RSGs, a novel population of 
   highly probable RSGs populating
   the more distant Scutum-Centaurus arm appears.}


   \maketitle

\section{Introduction}

Stars more massive than $\approx 8$ \Msun\ and below $\approx 40$ \Msun\
go through the red-supergiant (RSG) phase; this ranges from 8 to 25 \Msun\ 
when  rotation is included in the models 
\citep[e.g.,][]{limongi17,meynet00}.
RSGs are cool stars with effective temperatures (\Teff) 
below 4500 K and low gravity ($log(g) <  0.7$ cm s$^{-2}$), 
and they are metal-rich ($-0.5 < [Fe/H] < 0.5$ dex) and  luminous
($log(\frac{L}{L_\odot}) \ga 4$).  Their young ages, 
from 4.5 to 30 Myr, ensure that they are part of
the young  disk of spiral galaxies, as recently
observed in M31 and M32 \citep{massey21}.
A census of RSGs is fundamental for testing models
of stellar evolution and  core collapses.
RSGs are He-burning stars whose fate is governed by 
mass-loss and rotation. They are usually found  in large molecular 
complexes, 
which  populate the spiral arms of the Milky Way; 
only a small fraction of them (about 10\%) are
associated with stellar clusters 
\citep[e.g.,][]{messineo17,messineo19,smith15}.
However, a search for individually selected RSGs is 
more complex than in stellar clusters, because of dust obscuration and little 
knowledge of distances, and RSGs are often mistaken for
asymptotic giant branch stars (AGBs) because they have similar colors and magnitudes.
AGBs have different internal structures and nuclear reactions, 
as they have a degenerate core of CO and
burning fuel  in two concentric shells 
(H in the external shell and He in the inner shell). AGBs with masses below 8 \Msun\
trace  star formation that occurred from $\approx 40$ Myr to 13 Gyr ago,
and they have a much larger metallicity range. 
In the last decade, significant progress has been made in  distinguishing 
the  populations of RSGs and AGBs; this is thanks
to Gaia data, which provide the parallactic distances 
and optical stellar energy distribution (SED) of millions of stars, 
and the availability of photometric time series. 
In the coming years, spectroscopic data at a resolution 
higher than 10,000 will 
become available from the GALAH, Gaia DR4 RVS, 
and 4MOST surveys, providing metallicity for millions of stars
\citep[e.g., Introduction of][]{messineo21}.

The catalog of \citet{messineo19}\footnote{The catalog was revised
to include Gaia eDR3 parallaxes \citep{messineo21z,messineo23}.}
collected and analyzed stars 
previously reported in the literature 
as stars of K-M type and class I.
Unfortunately, a significant portion of the gathered 
late-type stars turned out to be faint giants, and the parallactic distances 
revealed that most of the classes collected from past literature 
are not reliable. 
There should be many thousands of RSGs in the Milky Way's grand spiral. 
There are 6400 RSGs in the spiral galaxies M31 and 2850 in M33 
\citep[][and references therein]{massey21b}. 
We are still unable to count and differentiate between the 
population of bright RSGs and the millions of fainter giants, 
making the census of obscured Galactic RSGs still unfeasible. 
According to \citet{gehrz89} estimates, the Milky Way should 
have at least 5200 RSGs.

In order to prepare for the upcoming spectroscopic era, 
lists of bona fide late-type stars with high luminosities  are required.
Following the division of the theoretical diagram of luminosity values versus 
temperatures in areas, \citet[Gaia DR2][]{messineo19} and 
\citet[Gaia DR3]{messineo23} kept those
 $\approx$400\footnote{The  extinction calculation determines 
 the exact number of stars in these regions.} 
 out of 1725 stars analyzed, located in areas A and 
B\footnote{As in \citet{messineo19}, Area A is defined as   \Mbol $< -7.1$ mag, 
which is the AGB limit. Area B is coded as  $-7.1 < \Mbol < -5.0$ mag
and $L/\Lsun > 51.3-13.33 \times log(\Teff)$, with  log(\Teff[K]) $> 3.548$.}, 
as highly probable RSGs.
\citet{messineo23} remarked  that before relying on the temperatures 
listed in the Gaia database for these brilliant cool stars, 
one should wait for the Gaia RVS spectra release.
\citet{healy24} report 638 candidate RSGs (cRSGs) in the Gaia DR3 catalog,
exceeding the number of bona fide stars of \citet{messineo19}. 
Indeed, by design, the catalog of Messineo et al.  did not consider new  
photometric cRSGs from Gaia. This number is 
still too low to account for the Galactic population.

Two main issues in the luminosity calculation exist.
Since mass loss has a significant impact on the brightest 
and coolest stars, the first problem is  the 
precise computation of extinction at the bright end side.
Second, because 8–9 \Msun\ stars might be in both evolutionary phases, 
there is a natural confusion at the faint side between RSGs and super-AGBs. 
A novel technique for estimating interstellar extinction 
for luminous late M-type stars is described by \citet{messineo24}.
Interestingly, the technique  uses near- and mid-infrared colors 
to predict interstellar extinction without taking 
into account the stellar environment.

Here, we examine  the determination of the extinction and luminosity 
of about 300 additional cRSGs (Sect. \ref{sec.selection}).
The envelope optical depth is empirically calibrated, 
and its effect on the bolometric corrections is measured.
In Sect. \ref{sec.distribution}, the XY distribution
of the new sample is shown to share the same features as  the
 bona fide   RSGs from \citet{messineo19}. 
A summary of the main findings is given in Sect. \ref{mysummary}.

\section{New cRSGs from Gaia-DR3 and 2MASS catalogs}
\label{sec.selection}
We preselected  2,167,423  data points with
the extinction-free magnitude \Ks$-1.311\times(H-K_{\rm S}-0.2)< 8$ mag
from the 2MASS catalog (470,992,970 entries) \citep{messineo24}.
Indeed, this range of extinction-free \Ks\ magnitudes 
encloses all RSGs in areas A\&B by \citet{messineo19}.

By using their 2MASS IDs we retrieved 1,941,750 Gaia DR3 matches
\citep[][]{gaiadr3}, of which
1,880,040 were assigned distances in the catalog by \citet{bailer21}.
However, only 851,612 of those  have 
$\frac{\varpi}{\sigma_\varpi (est)}$ larger than 4, where
$\varpi$ is the parallax and $\sigma_\varpi (est)$
is the external parallactic error \citep{mais21}.
850,926 data points have 2MASS $J$, $H$, and \Ks\ data (all three bands). 

\begin{figure}
\begin{center}
\resizebox{0.99\hsize}{!}{\includegraphics[angle=0]{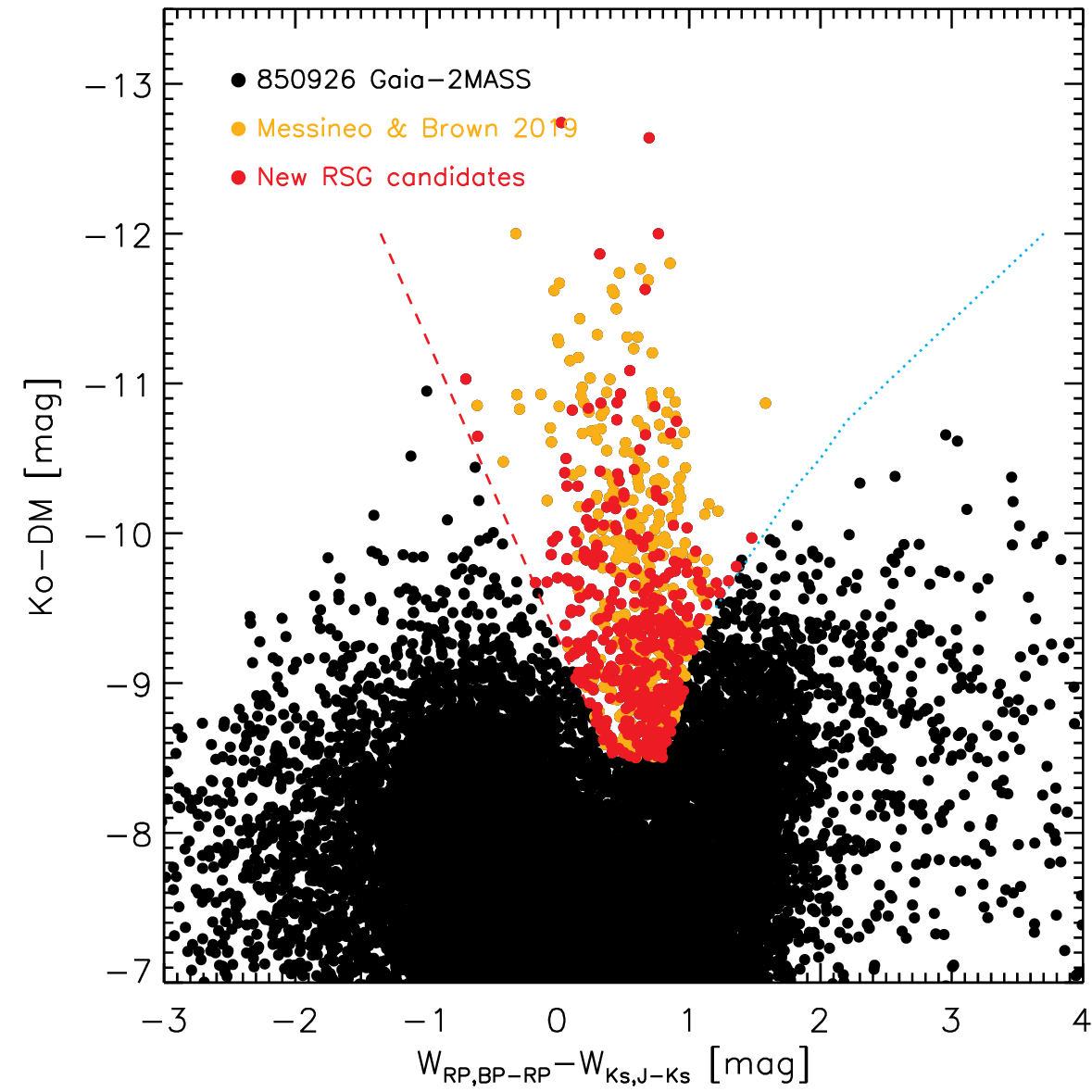}}
\end{center}
\caption{ \label{fig_figueras} \Mk\  versus 
W$_{\rm RP,BP-RP}-$W$_{\rm K_S,J-K_S}$ $<4$ colors of 2MASS stars brighter than 
\Ks$-1.311 \times (H - $\Ks$ - 0.2) < 8$ mag and with good Gaia parallax. 
Selected cRSGs (in red and orange) lie in the cone 
enclosed by the two curves. The dotted cyan curve indicates a rough separation 
between  O-rich AGB stars and C-rich AGB stars (to the right).
The long-dashed red curve separates RSGs from O-rich AGBs and S-type stars.
Stars included in the catalog of \citet{messineo19} are colored in orange.}  
\end{figure}

Following the diagnostics described in \citet{messineo23},
the diagram of the absolute \Ks\ values, \Mk, versus the extinction-free
color  W$_{\rm RP,BP-RP}-$W$_{\rm K_S,J-K_S}$, calculated as in \citet{abia22}, 
was analyzed. RSGs appear to be preferentially located in a narrow range 
of colors between 0 and 1 mag,
while the O-rich AGB stars are bluer than that, and C-rich stars are redder 
than that. In this diagram, two curves are drawn to roughly indicate  
two separation lines between C-rich and RSGs, and between O-rich and RSGs. 
730 cRSGs with  \Mk $< -8.5$ mag 
are located in the cone enclosed by these two curves,  
as illustrated in Fig.  \ref{fig_figueras}. 
As 315 cRSGs are already included in the list of \citet{messineo19},
415  stars remain. 
We searched for available spectral types of the  415 cRSGs 
in the catalog by \citet{skiff16}
and in SIMBAD and retrieved it for 279 stars --
185 stars are M-type stars, 22 K-type stars, 20 C-rich stars, 
32 S-type stars, 14 F-G-types, and  seven early-type stars.
The B-A-F-G type, C-rich, and S-type stars were dropped 
from the sample, which remains composed of 342 O-rich stars, of 
which 207 with known KM spectral types and 135  previously unclassified.
As shown in the following sections, additional
spectral types were estimated with Gaia BPRP spectra bringing 
the number of available spectral types to 335,
and  282 those in areas A and B.

Recently, \citet{healy24}  published a
catalog of chosen RSGs from the Gaia DR3 database,
which comprised 638 entries. 
We count 788 stars when we concatenate the stars from areas 
A and B of \citet{messineo19}, 
\citet{messineo23}, and from the current work.
The Messineo's and Healy et al.'s compilations share 490 entries.
Only 12 cRSGs from the selection presented in this work
are included in Healy's catalog. 
This motivated us to publish this additional list of cRSGs
which may be useful to plan forthcoming and ongoing
spectroscopic surveys 
\citep[e.g., GALAH, 4MOST;][]{sharma22,dejong12}.

\subsection{Available photometry}
\label{sect.phot}
The stars were selected from Gaia  and 2MASS catalogs. 
The Two Micron All Sky Survey (2MASS)  survey is described in \citet{skrutskie06}, and
in \citet{cutri03}, and in the online 
manual\footnote{\url{https://irsa.ipac.caltech.edu/data/2MASS}}.
The stars are extremely bright at infrared wavelength, and 
all measurements come from the 2MASS short-exposure frames (51ms).
The $J$-band measurements range from 1.89 to 8.68 mag, with a peak around 5 mag,
while the $H$-band magnitudes range from 0.96 to 14.77 mag  and peak at 3.5 mag,
and the \Ks\ magnitudes range from 0.66 to 7.70 mag with a peak at 3.0 mag.
Stars that were saturated in the short integration frames had their 
profile reconstructed using the unsaturated pixels.
While magnitudes of unsaturated point sources (fl\_red=1)  have a 
typical uncertainty of 0.01-0.02 mag,
magnitudes of saturated sources  (fl\_red=3) are accurate within 0.2 mag.
For  5\%  of the stars  measurements  from the  
Catalog of Infrared Observations (CIO) 5th edition by
\citet{gezari96} are available; $K$ measurements are available for all 
stars in the subsample:
$H$ mag for 44\% and $J$ mag for 22 \%.
The average difference between the  CIO and  2MASS $J$  
magnitudes brighter than 4 are $+0.15$ with a standard deviation $\sigma$=0.12 mag.
Between the CIO  and  2MASS $H$  magnitudes brighter than 4, it is 
$+0.24,$ with a standard deviation $\sigma$=0.44 mag; 
the average difference between the CIO and the 2MASS \Ks\ 
magnitudes brighter than 4 is $+0.00$ mag, with a standard deviation of $\sigma$=0.62 mag;
however, after excluding J03232718+6545417 and J08104310$-$1519469, 
the average difference becomes 0.07 mag with a $\sigma$ of 0.17 mag.

Positional matches (within 5\arcsec) in the Midcourse Space Experiment 
catalog \citep[MSX,][]{egan03}  were found for 89\% of 
the newly selected stars; 94\% were found in the 
Wide-field Infrared Survey Explorer (WISE) survey \citep{wright10}, 
25\% were found in the Galactic Legacy Infrared Midplane Survey 
Extraordinaire (GLIMPSE) survey \citep{churchwell09}, 
and 26\% were found in the MIPSGAL 24  \um\ survey 
\citep{gutermuth15} using a search radius of 4\arcsec.

$JH$\Ks\ measurements were all retained including 2MASS upper limits; however,
a flag is given to indicate sources with the good 2MASS photometry.
The SED of each star was visually inspected
and a few measurements which appeared unrelated to the SED were removed.
2MASS J20434317+4741398, J17594120-2857418, and
J14090677-5943243   had their $H$-band upper-limit magnitudes removed,
2MASS J18095816-2728222 had the WISE measurements removed (confusion and blend),
2MASS J18440043-0457062 had the WISE W3  and W4 magnitudes removed (confusion),
2MASS J10572371-6046304 had the WISE W4 upper limit magnitude removed,
and 2MASS J19283960+1718220 had its GLIMPSE [3.6] magnitude removed (artifact on image).

\begin{figure*}
\begin{center}
\resizebox{0.48\hsize}{!}{\includegraphics[angle=0]{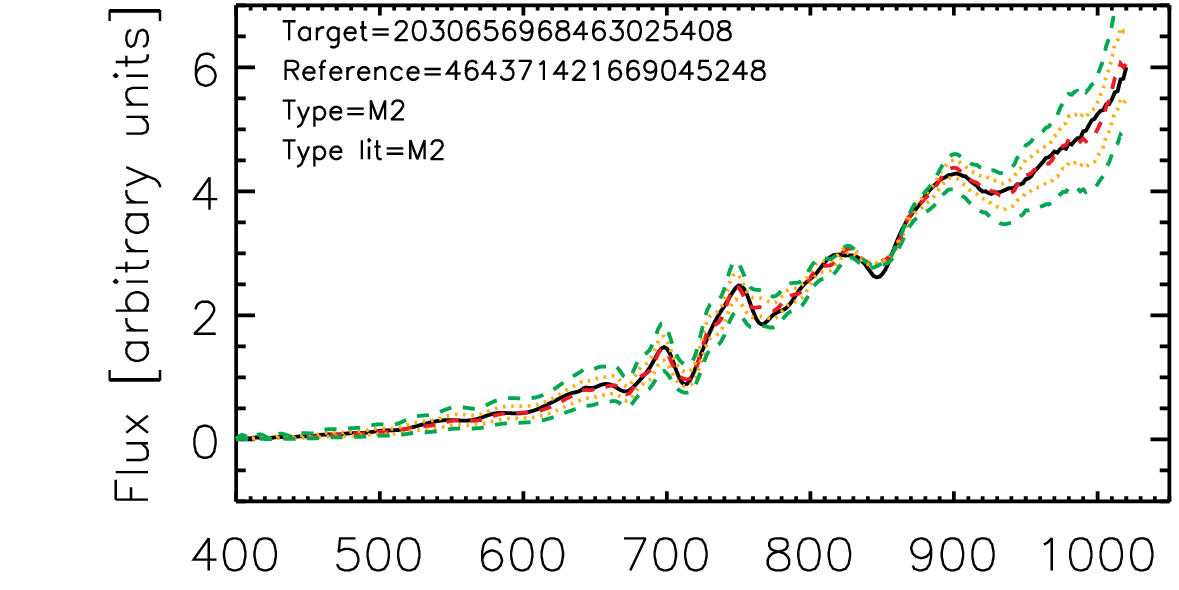}}
\resizebox{0.48\hsize}{!}{\includegraphics[angle=0]{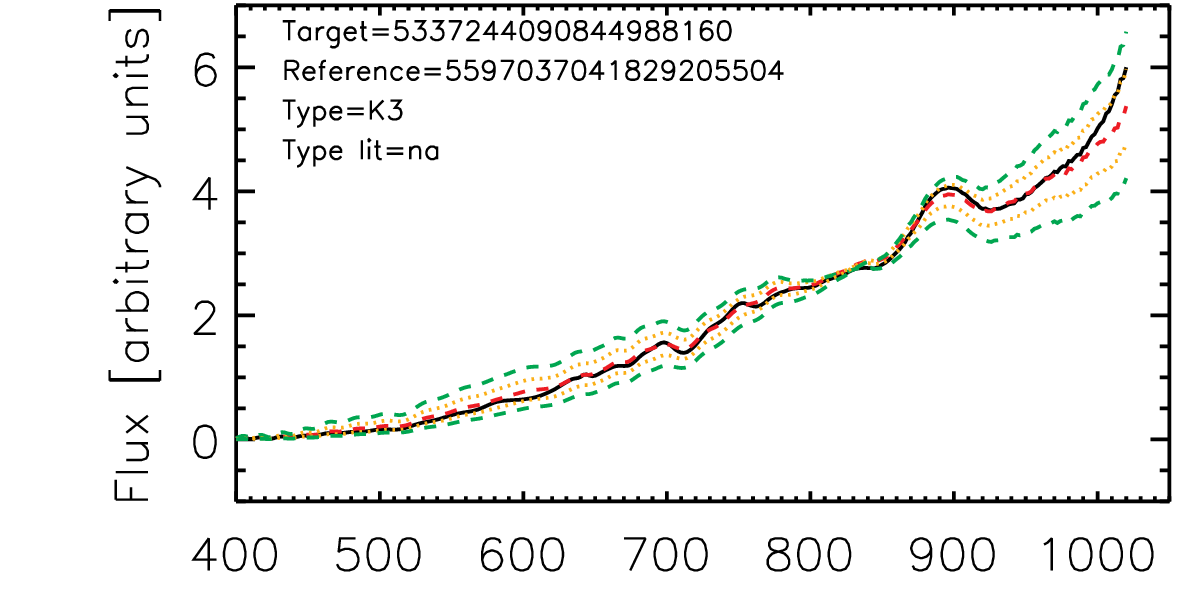}}
\end{center}
\caption{ \label{fig_deltaext} 
Two examples of BPRP spectra. The target spectrum is shown with a black
curve, and the reference spectrum  with the dashed red curve.
The reference spectrum  was brought to the target's extinction, 
which is estimated in the infrared.
Extinction variations of $\Delta$\Aks$=\pm0.05$ mag and $\pm0.10$
are indicated with orange dotted  and  green dashed curves, 
respectively. Gaia source\_ids are given in the figure labels.}  
\end{figure*}

\subsection{Spectral types,  temperatures, and extinction  of cRSGs}
\label{teffscale}

In this work, a selection of cRSGs is presented. 
Although estimates of stellar parameters are provided, 
they are primarily used to ensure a good selection of stars. 
More accurate determinations of stellar parameters will require 
high-resolution spectroscopic follow-up observations and updated photometry.

The determination of \Teff\ in RSGs is a challenging task. 
A direct determination of \Teff\ requires the knowledge of the stellar radii, 
which are only available for a limited number of stars through interferometric measurements.
More commonly, \Teff\ values are 
estimated using color temperatures or spectroscopic temperatures
 \citep[e.g.,][]{levesque05}. 
Historically, the TiO bands visible in the optical spectra of RSGs 
have been used to determine their spectral types.  By using
models of  stellar atmospheres (such as the MARCS models) 
and a fitting technique, \citet{levesque05}
established a TiO-based temperature scale.
However, as discussed in \citet{davies13}, these models are 
far from perfectly reproducing observed spectra, and there is a 
degeneracy between extinction and temperature determination. 
Furthermore, both \citet{davies13} and \citet{levesque18}
demonstrated that the 1D MARCS models fail to accurately 
reproduce infrared fluxes, and using either the optical or 
infrared portion of the SED 
can result in different \Teff\ values 
(and, consequently, extinction estimates).

These findings raise concerns about the use of TiO bands 
for determining \Teff, leading to alternative attempts at 
temperature determination using atomic lines 
\citep[see, e.g.,][]{gazak14, tabernero18, dicenzo19,taniguchi21,taniguchi25}.
\citet{messineo21} show that the temperature measurements from 
\citet[][]{gazak14} based on $J$-band spectra (R = 11,000-14,000)
agree with those of \citet{levesque05} to within 150 K.
\citet{tabernero18} propose a new temperature scale 
based on spectral features in the CaT range (R = 7,000), 
confirming a correlation between TiO band depths and stellar temperatures. 
\citet{taniguchi25} determine RSG temperatures using 
YJ spectra (R = 28,000) and 11 pairs of atomic iron lines; 
their temperatures are in good agreement with the
\Teff\ from TiO of \citet{levesque05} to within 100 K.

In conclusion, the use of TiO bands to determine temperatures 
remains valid, as it is supported by studies of atomic lines. 
Currently, spectroscopic temperatures can be used as a proxy for \Teff, 
with a typical accuracy level of 100–150 K.

\subsubsection{Spectral types}
We analyzed the BPRP spectra to 
estimate KM types of the newly selected 342 O-rich stars.
327 BPRP spectra were available, 
but four were excluded because of poor quality,
and four stars had spectra typical  of G-types.
Among the 15 stars without BPRP spectra, three do not have  types in SIMBAD. 
In conclusion, a clean sample of 335 bright K- and M-type stars 
was produced.

\begin{figure*}
\begin{center}
\resizebox{0.99\hsize}{!}{\includegraphics[angle=0]{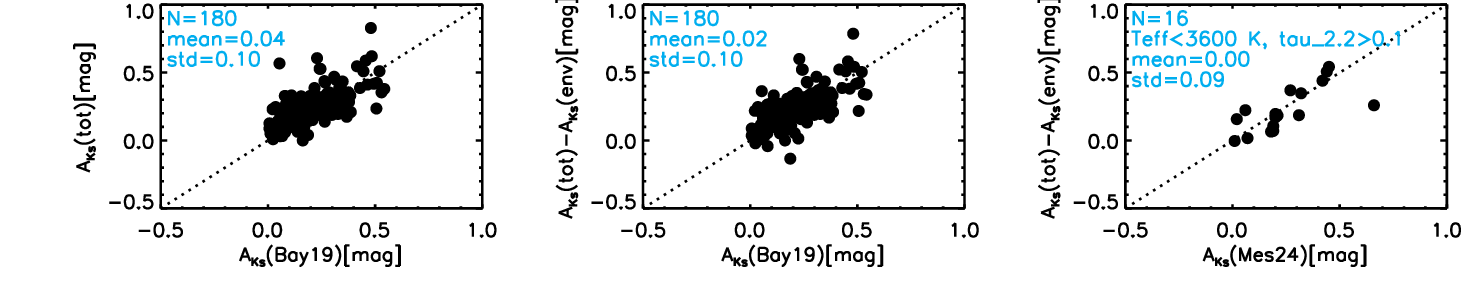}}
\end{center}
\caption{\label{bay19} {\it Left panel:} 
Estimated \Aks(tot) values versus 3D dust map \Aks(Bay19) values.  
{\it Center panel:} \Aks(tot)-\Aks(env) values versus 3D dust map \Aks(Bay19) values. 
{\it Right panel:} For 16 stars with \Teff$ < 3600$ K and $\tau_{2.2} > 0.1$, the 
\Aks(tot)$-$\Aks(env) values versus  \Aks(Mes24) values are plotted. 
}  
\end{figure*}

The sample contains 135 previously unknown  late-type stars, which
we classified solely using the Gaia DR3 BPRP spectra.
Because of  a degeneracy between temperatures and reddening, 
an independent initial guess of interstellar extinction was required.
Late-type stars span a small range of  naked $H-K_S$ colors, 
and, initially, we assumed 
a fixed M1-type star with a naked  $H-Ks=0.22$ mag 
\citep{koornneef83} and an interstellar power law with an index of $-2.1$
\citep{messineo05,messineo24}; because the associated color uncertainty
is typically below  0.12 mag (the delta color between an M1 and an M5),
we estimated an \Aks\ uncertainty  within 0.16 mag.
The optical extinction curve by \citet{cardelli89}, extrapolated 
to the near-infrared with a power law of index $-2.1$, 
was used to de-redden the BPRP spectra.
Spectral types were inferred by matching 
the de-reddened BPRP spectra of the target stars with those  
of reference stars, as described in \citet{messineo23}. 
A few types were visually inferred if the solution diverged.
The process was reiterated;
with the obtained spectral types, temperatures were estimated using
the temperature scale of \citet{levesque05},
 and the total \Aks\ 
extinction values were recalculated  with the color tables of \citet{koornneef83}.
Then, spectral types were re-estimated with the new total \Aks\
and the spectral library. On the second iteration, as spectral types
had been estimated, the total \Aks\ values were estimated from 
the $J-$\Ks\ color excesses. 
The larger separation in wavelengths of the two filters
ensures a more sensitive meter. Furthermore, the stars
are very bright, and, generally, the
fainter $J$ magnitudes are of better quality than the $H$ 
and \Ks.  The total \Aks\ estimates from $H$\Ks\  have an average 
error of 0.34 mag, while the total \Aks\ estimates from $J$\Ks\ have an average error of 0.11 mag.
In Fig. \ref{fig_deltaext}, two examples of BPRP spectra are shown
along with the selected reference spectrum at the target's
extinction. 
A comparison with the 3D dust map of \citet{green19} 
and our total \Aks\ values was possible for 180 stars; 
after transforming the  map E(B-V) values to 
\Aks(Bay19)$=E(B-V) \times 3.1 \times 0.077$ mag,
a mean difference of 0.04 mag was measured with a sigma of 0.10 mag,
as shown in Fig. \ref{bay19}.
However, 3D Galactic extinction maps are made with a spatial 
resolution of several arcminutes; while our total \Aks\ values  were estimated 
with the infrared colors of the considered star.
We retained our calculation of extinction. 
As explained in the following section, for most of the stars,
the interstellar extinction \Aks(int) can be approximated with 
the total stellar extinction \Aks(tot) because the envelope 
optical depth is negligible.

\begin{figure}
\begin{center}
\resizebox{0.99\hsize}{!}{\includegraphics[angle=0]{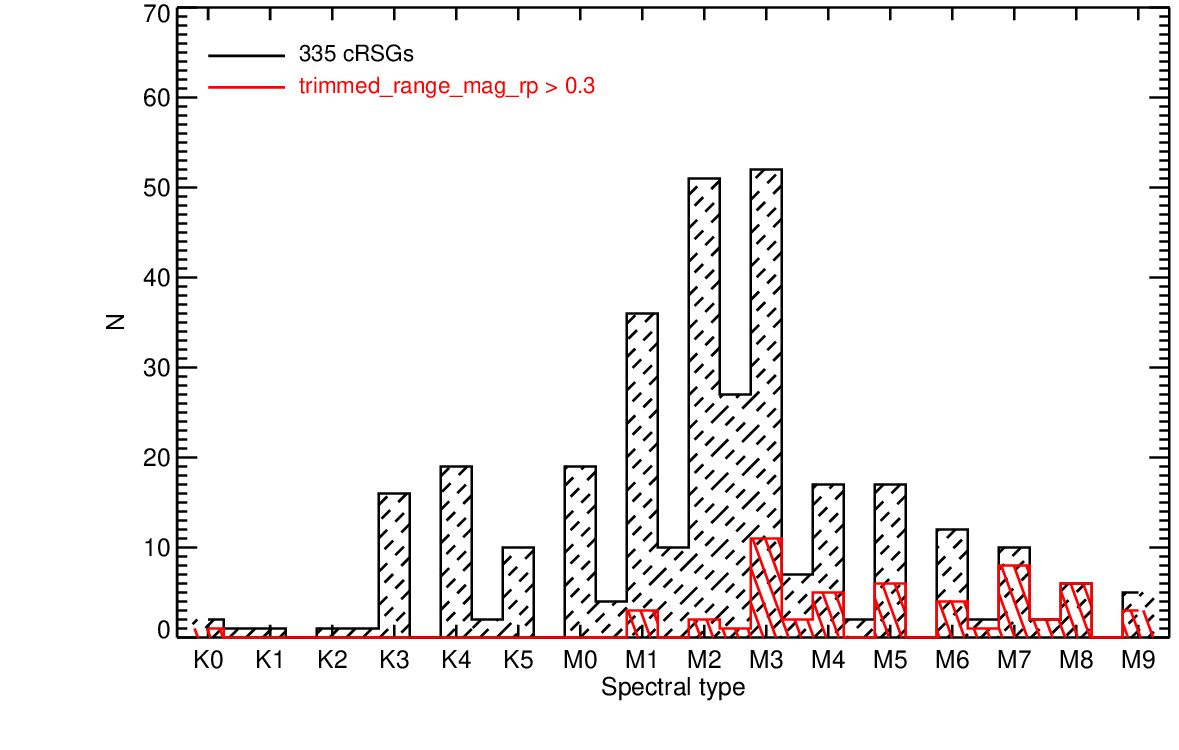}}
\end{center}
\caption{ \label{histo_sp} Distribution of spectral 
types of 335 late-type stars analyzed. The histogram of 
stars with trimmed\_range\_mag\_g\_fov $< 0.5$ mag is shown in red.}  
\end{figure}

Using the subsample of stars with spectral types in the literature,
it was estimated that the accuracy of the inferred spectral types mostly  
fit within two spectral types.
In Fig. \ref{histo_sp}, the distribution of spectral types appears to peak
at M2-M3 types, as expected for RSGs \citep[see Fig. 5 in][]{messineo19}.
The newly selected sample of 335 $K$- and $M$-type stars still includes
variable stars. The histogram of stars with a  trimmed\_range\_mag\_g\_fov 
larger than 0.3 mag is over-plotted in red. 
The variables are the dominant population for M5-M9 types.
When only considering stars in areas A or B of \citet{messineo19}, 
the final sample consists of 282 stars.

\subsubsection{Envelop extinction}
We assumed that the interstellar extinction  \Aks(int) 
can be accurately approximated by the total extinction  \Aks(tot)
computed using the intrinsic colors  of naked stars.
The stars are by design  the reddest in the 
W$_{\rm RP,BP-RP}-$W$_{\rm K_S,J-K_S}$ color, as expected for 
early-M types, 
and their mass-loss rate is anticipated to be mild. 
The envelope optical depth of RSGs typically falls 
between 0.0 and 0.7 at 0.55 \um\ \citep{humphreys20}.

The optical depth of circumstellar envelopes can be measured 
by fitting infrared flux densities with DUSTY models. 
The mid-infrared features observed in the de-reddened stellar 
SED are intrinsically linked to the stellar envelope. 
Effective extinction ratios at mid-infrared wavelengths 
typically range from 0.3 to 0.5 times the value of \Aks\ 
\citep{messineo24}. 
These small variations in effective extinction ratios imply that, 
during the de-reddening process for interstellar extinction, 
the characteristics of these features remain consistent; 
an emission feature will consistently appear in emission, 
indicating an optically thin envelope. 
In summary, the envelope $\tau$  at 2.2 \um\ 
is estimated by examining the presence and strength of the 
silicate feature at 10 \um\ in the SED, alongside the flux 
excess at wavelengths longer than 8 \um.

\begin{figure}
\begin{center}
\resizebox{0.7\hsize}{!}{\includegraphics[angle=0]{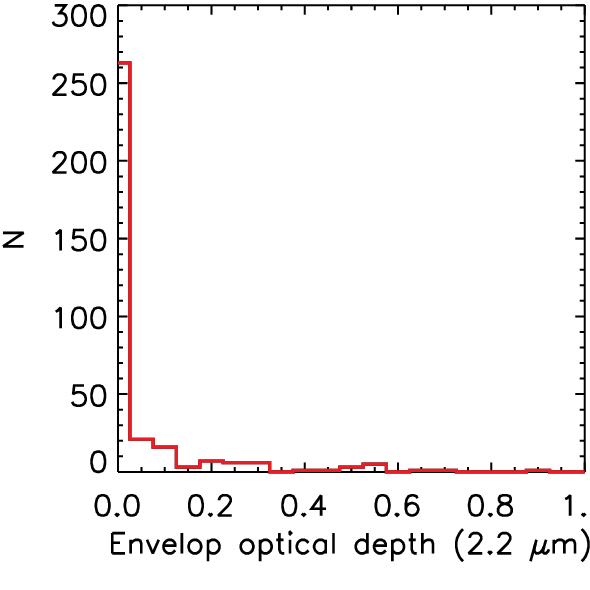}}
\end{center}
\caption{ \label{histo_tau} Distribution of envelope optical depth
at 2.2 \um\ of the newly selected late-type stars. }  
\end{figure}

A grid of SED models was 
constructed using the DUSTY algorithm of 
\citet{ivezic99}.
The NextGen spectra of \citet{allard11} with 
log$_{10}$(g)=0.5, solar metallicity [M/H]=0 dex, and \Teff\ from 2600 to 4500 K
in steps of 100 K were utilized as input stellar models\footnote{ The spectra 
are distributed by the Virtual Observatory SED analyzer
\citep{bayo08}. The bt-nextgen\_agss2009 (gas only) were retrieved.}. 
Cold silicates by \citet{suh99}
were used, and the  optical depth at 2.2 \um, $\tau_{2.2}$,
was varied in increments of 0.02 
from 0 to 1.5.  A maximum size of 1.000 \um\ and a 
minimum size of 0.001 \um\ were adopted for the 
Mathis–Rumpl–Nordsieck (MRN) size distribution. 
The most basic spherical envelope assumption was made
with  the density decreased as R$^{-2}$ 
and a dust condensation radius at 1000 K.
The temperature scale of \citet{levesque05}
was used to translate the stellar spectral types 
(see the section above) to temperatures.
After selecting the set of models that were closest to the 
stellar temperature, the $\tau_{2.2}$ was deduced 
by minimizing the discrepancies
between the infrared observed fluxes  (F$_\lambda$) 
and the model flux densities at the corresponding  isophotal wavelengths.  
The theoretical SEDs 
obtained from synthetic spectra and the DUSTY radiative code 
were compared with the observed SEDs. 
When varying the stellar \Teff\ of $\pm100$ K, 
the solution converged to the same $\tau_{2.2}$.
While the majority of the observed SEDs are consistent with models of naked stars, 
some exhibit an emission feature around 10 \um\, 
accompanied by a flux excess beyond 8 \um 
\footnote{This simple approach (SED modeling) 
could not be used for AGBs by 
\citet{messineo24} and \citet{messineo18} because the AGB 
Miras are strongly variable;  
by using non-simultaneously taken single-epoch measurements, their SEDs
appeared as a zigzag without displaying a clear 10 \um\ feature.}.
Precisely, as seen in Fig.\ \ref{histo_tau}, 
the bulk of stars have negligible circumstellar extinction  
with $\tau_{2.2} < 0.02$; for 7\% of the sample, 
$\tau_{2.2}$ is larger than 0.25 (from 
0.25 to 0.90), their SEDs show visible silicate features in emission, and  
a more accurate evaluation of interstellar extinction is needed.

The obtained $\tau_{2.2}$ is 
useful for relative comparisons of different envelope types. 
To compare this theoretical \(\tau\) with the  
\Aks(int) and \Aks(tot) inferred from observations, 
we needed to scale the model \(\tau\) to match the observed 
\Aks(tot)$-$\Aks(int).  \Aks(tot) is an 
equivalent total extinction because it is the total
extinction obtained using the interstellar extinction law. 
The scaling parameters will depend 
on various factors, including the adopted grain composition 
(material and sizes), envelope geometry, and dust temperature. 
In this study, we assumed that the silicate models proposed by 
\citet{suh99} are appropriate for the envelopes of  cool 
cRSGs and will primarily explore the effect 
of the maximum grain size on the resulting \(\tau\).\\
An estimate of interstellar extinction, \Aks(Bay19), was obtained from
maps \citep{green19}. \\
Another independent estimate of interstellar extinction, 
\Aks(Mes24),  was obtained with the technique outlined  by
\citet{messineo24} for O-rich Mira stars.
For mass-losing Miras, a fiducial sequence of intrinsic  color-color 
points is adopted (e.g.\ in the plane ($J-$\Ks$)_o$ vs. (\Ks$-[24])_o$), 
and used, along with the extinction-free color definition, to infer
the interstellar extinction.  
For example, Q$_{24}^{JKs}$=$J-$\Ks$-{\rm const} \times $\Ks$-[24]$=
$(J-$\Ks$)_o-{\rm const} \times ($\Ks$-[24])_o$).
By measuring Q$_{24}^{JKs}$, ($J-$\Ks$)_o$ and (\Ks$-[24])_o$ 
can be estimated; from the de-reddened  ($J-$\Ks$)_o$ and (\Ks$-[24])_o,$ 
the interstellar \Aks(Mes24) is estimated, 
as explained in \citet[][]{messineo24}.\\
For 16 stars with DUSTY $\tau_{2.2} > 0.1$ and \Teff\ below 3600 K, 
budgeting of the interstellar extinction \Aks(Bay19) and \Aks(Mes24), and 
 the total extinction \Aks(tot)  was made.
The \Aks(int)=\Aks(tot)-$\tau_{2.2}\times {\it factor},$ and the $factor$ was chosen
to avoid negative \Aks(int)\footnote{
DUSTY $\tau_{2.2}$ decreases with decreasing maximum grain size.
$\tau_{2.2}$[0.001-1.000]=$2.78\times \tau_{2.2}$[0.001-0.350]
and
$\tau_{2.2}$[0.001-1.000]=$3.40\times \tau_{2.2}$[0.001-0.250].
}.
The  \Aks(int) values
 must be in agreement with the \Aks(Mes24) and \Aks(Bay19);
we obtained that \Aks(int)$\approx$\Aks(tot)-$\tau_{2.2}\times0.3$ 
(see Fig. \ref{bay19}), and  the envelope extinction 
\Aks(env)$\approx 0.3\times\tau_{2.2}$[0.001-1.000].
When the maximum gran size is 0.25-0.35 \um,  
the $factor$ becomes unity; i.e., interstellar and circumstellar extinction curves
become similar. 

When neglecting the circumstellar envelope and assuming  \Aks(tot)
as  \Ak(int),
the maximum overestimate of \Ak(int) would amount 
to 0.27 mag for the new sample.
The correction \Ak(int)=\Aks(tot)$-$\Aks(env) 
was applied and the \Mbol\ recalculated.

\subsection{Luminosities and bolometric corrections}
\begin{figure}
\begin{center}
\resizebox{0.99\hsize}{!}{\includegraphics[angle=0]{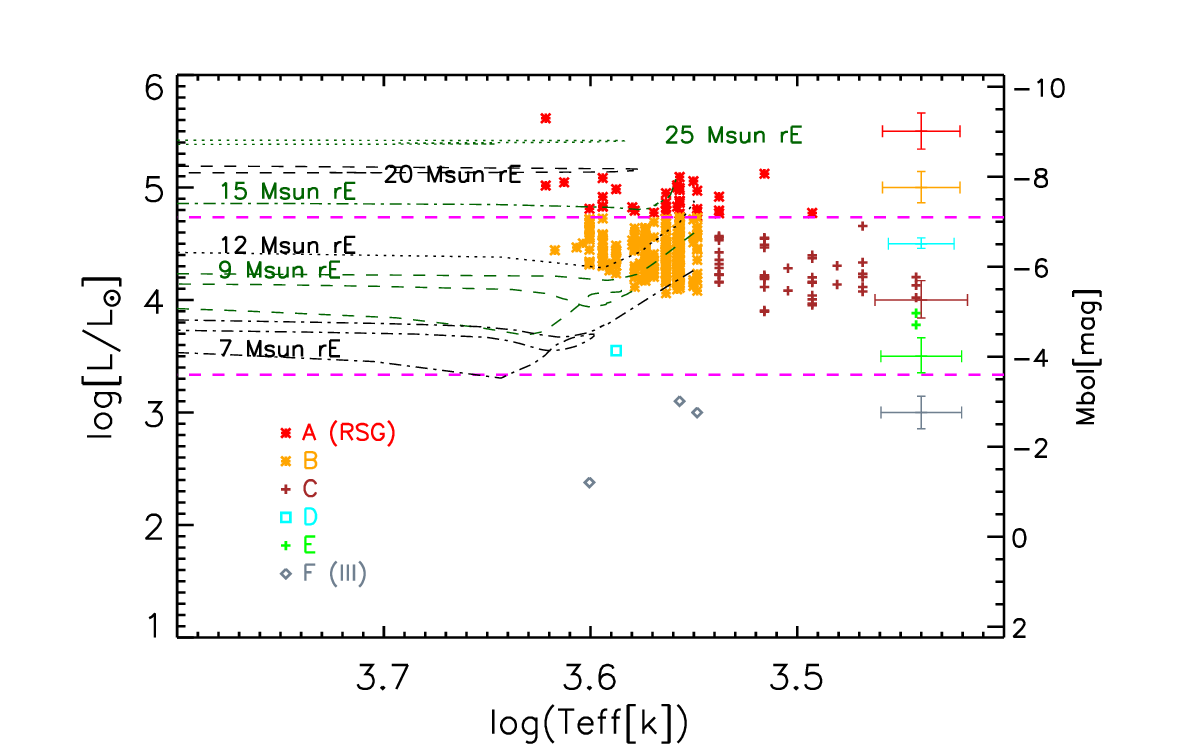}}
\end{center}
\caption{ \label{fig_lum} Luminosities versus \Teff\ of newly selected stars. 
 Evolutionary tracks with solar metallicity and rotation by \citet{ekstrom12}
are over-plotted.}
\end{figure}

\begin{figure}
\begin{center}
\resizebox{1\hsize}{!}{\includegraphics[angle=0]{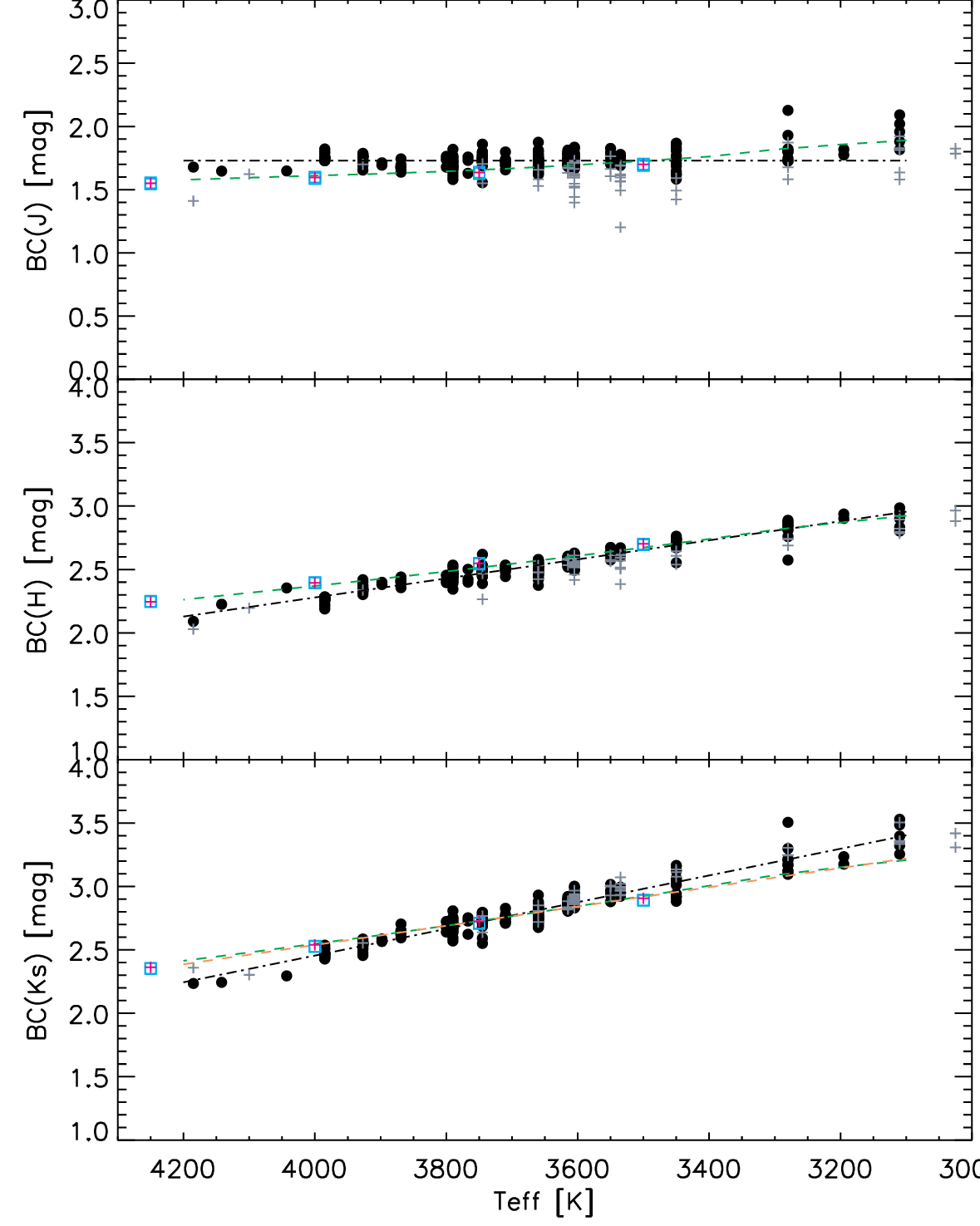}}
\end{center}
\caption{ \label{fig_bc} {\it Top panel:} For stars with $\tau_{2.2} < 0.08$, bolometric corrections BC$_{\rm J}$ 
are plotted versus \Teff\ values with  black-filled circles. 
For stars with $\tau_{2.2} > 0.08$, gray crosses are plotted.
The BC values were calculated with the \Mbol\ inferred with the atmosphere 
models by \citet{allard11} and the DUSTY code 
and the absolute $J$ magnitudes (from 2MASS).
The dotted-dashed black line is the best polynomial 
fit of degree zero  to the data points.  
The dashed green line is the theoretical BC$_{\rm J}$ inferred with 
Allard's model ([M/H]=0, log(g)=0.5).
The cyan squares mark the BC$_{\rm J}$
values estimated with an ATLAS9 grid of models and 
for Bessel filters by \citet{howarth11}
([M/H]=0.0 and log(g)= 0.5 dex); while our re-calculations for the
2MASS filters are marked with red crosses.
{\it Middle panel:} Bolometric corrections BC$_{\rm H}$ 
versus \Teff\ values.
Lines and symbols are as for the top panel.
{\it Bottom panel:}  
Bolometric corrections BC$_{\rm Ks}$ 
are plotted versus the \Teff\ values.
Lines and symbols are as for the top panel.
The dashed-orange line is the $K$-band 
relation for naked RSGs found with MARCS 
models  and the $K$ -filter profile of 
\citet{bessell88} by \citet{levesque05}.
}
\end{figure}

\begin{figure}
\begin{center}
\resizebox{0.99\hsize}{!}{\includegraphics[angle=0]{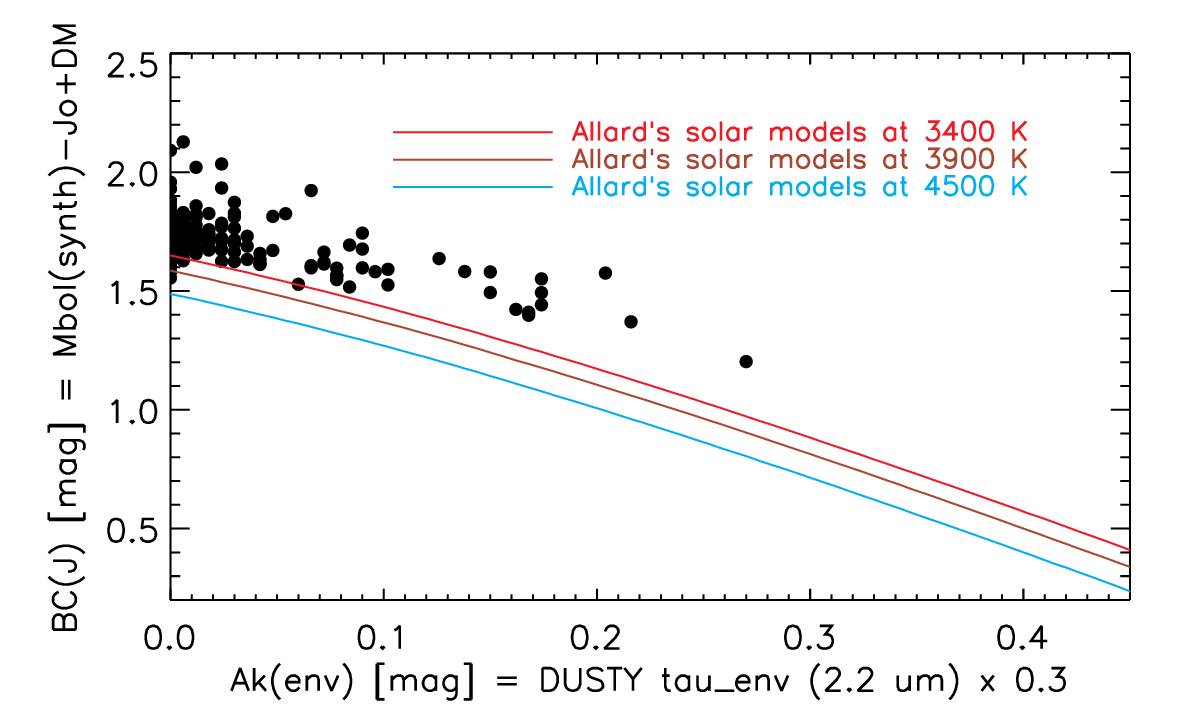}}
\end{center}
\caption{ \label{bctau} Empirical BC$_{\rm J}$ values  
$(\rm \Mbol(DUSTY)-(2MASS.J-3.208\times A_{\rm K_{\rm s}}(int)- DM))$
are plotted versus inferred \Ak(env.). 
For supergiants (log(g)=0.5) of solar metallicity and \Teff\ of 3200, 3900, and
4500 K, the theoretical BC$_{\rm J}$ calculated with  the 
DUSTY simulation and  Allard's models are over-plotted.}
\end{figure}

One set of luminosities was obtained using the de-reddened 2MASS \Ks, 
the Gaia distances of \citet{bailer21},
the  temperature scale, and the \BCK\ of \citet{levesque05}.
The infrared extinction curve was approximated with a power 
law of index $-2.1$ \citep{messineo24}.
When changing the assumed spectral type of one unit 
(from M3 to M2 or M4),
\Aks\ varies by 0.03 mag on average ($\sigma$=0.01 mag), and
the bolometric magnitudes \Mbol(BC$_{\rm Ks}$) change by $0.09$ with sigma=0.03 mag.

A second set of luminosities was derived with a trapezoidal 
integration under the SED made
with the available infrared flux densities 
and by extrapolating to the blue side
with the blackbody of the same temperature as the  star, 
and to long wavelengths 
with a linear extrapolation to zero flux at 500 \um,
as in \citet{messineo19}.  
The trapezoidal integration depends on the temperature 
through the blue extrapolation and extinction calculation.
When changing the assumed spectral type of one unit (from M3 to M2 or M4),
after recalculating \Aks, the trapezoidal magnitudes change to $0.09$ with sigma=0.03 mag.

A third estimate of bolometric magnitudes (\Mbol(DUSTY)) 
was obtained  by integrating under the DUSTY model 
that best fit the observed infrared flux densities (de-reddened).
The model was scaled so that the average of the observed 
$JHKs$ flux densities best matched the average of the model 
flux densities at the isophotal wavelengths of the three 2MASS filters.
The integrations were obtained in the plane F$_\nu$ d$\nu$, the
adopted zero-point constant was assumed to be $-18.997$, and 
the solar \Mbol= +4.74 mag \citep[][]{mamajek15}.
The bin used for the $\tau_{2.2}$ grid corresponds to
an \Mbol\ uncertainty  $Mtau_{err} < 0.01$ mag, and the
$\tau_{2.2}$ value remains unaffected by \Teff\  
variations  within 100 K in the model.
By shifting the best models of $\pm100$ K,  
\Mbol\ variations are  within 0.07 mag. 
When changing the assumed spectral type of one unit,
after recalculating \Teff\ and \Aks, 
the \Mbol(DUSTY) changes by $0.07$ with sigma=0.03 mag.

The dominant errors in \Mbol\
come from the photometric uncertainties 
($\pm 0.14$ mag, when  simultaneously decreasing 
or increasing the magnitudes by their errors) 
and the distance moduli
(median errors are [+0.24,$-$0.27] mag, 
average errors are [+0.24,$-$0.27] with $\sigma$ of [0.09,0.12] mag,
see Table 1).

To calculate the global statistical errors on the three \Mbol\ values, 
a Monte Carlo simulation was carried out (with 2,001 runs).
For each star, a table of 2,001 Gaussian random errors was extracted 
for each available magnitude  (using a different seed for each); 
each Gaussian had a $3\times\sigma$ equal to the quoted magnitude error.
For each adopted spectral type, a Gaussian error with $\sigma$ equal to 
two spectral types was randomly added.
For each random extraction,  magnitudes, spectral types, \Teff,
\Aks, \Mbol(BC), \Mbol(trap), and \Mbol(DUSTY)  values
were recalculated. The average  magnitudes and $\sigma$ values
are given in Table 1.

The average  difference between the
\Mbol\ values from the \BCKs\ of \citet{levesque05} 
and those from the DUSTY models
 is $-0.03$ mag, with $\sigma=0.09$ mag.
The average  difference between the
trapezoidal \Mbol\ values and those from the DUSTY model
is $-0.23$ mag, with  $\sigma=0.08$ mag.

For the bulk of stars, the derived luminosities are consistent with those
of RSGs, as shown in Fig. \ref{fig_lum}.
They are mostly located in areas A and B of the luminosity-versus-temperature diagram.

The theoretical \BCJ, \BCH, and \BCKs\ values estimated with the NextGen models 
([M/H]=0, log(g)=0.5) of \citet{allard11} are listed 
in Table \ref{allardBC}.
The stellar temperature uncertainty affects the luminosity and 
the bolometric correction values.
For the BC values of naked stars, uncertainties on the BC values 
depend on the slope of the relation between the BC  values
and  the stellar \Teff\ values, as explained  by \citet{levesque18}
and shown in Fig. \ref{fig_bc}. 
The predicted slope in the \Ks\ filter by  Allard's models
is almost identical to that predicted with MARCS models by \citet{levesque05}.
The $J$ band is particularly useful, because the gradient of the BC 
with increasing stellar temperature is negligible (for \Teff$>3400 $ K).
Differences in \Teff\ corresponding 
to one nearby spectral type (e.g., from M0 to M1) amount to  45--85 K
\citep{levesque05}. Therefore, 
a typical uncertainty in the \Teff\ scale  of 100--150 K 
(as noted in recent literature and described in Sect.\ \ref{teffscale}) 
corresponds to two spectral subtypes (e.g., from M1 to M3). 
In the $J$ band, for \Teff\ from 3400 to 4500 K, an uncertainty of 100 K 
implies an uncertainty in  \BCJ\ of 0.015--0.035 mag, while
in the \Ks\ band, it implies an uncertainty in \BCKs\ from 0.063 to  0.082 mag.
However, the selection of the band must also take into account 
the effect of circumstellar extinction, which 
decreases with increasing wavelengths. 

The presence of  thick circumstellar envelopes 
 changes the BC values, which can    significantly deviate from those  
 for naked stars 
\citep{messineo04,davies13,messineo24}.
To understand the impact of   envelope extinction on the
\Mbol\ values, the  empirical bolometric correction in bands $J$, $H$, and 
\Ks, \BCJ, \BCH, and \BCKs\ values, were calculated as follows:
$${\rm \BCJ= \Mbol(DUSTY)-(J-3.208\times A_{\rm K_{\rm s}}(int)- DM)\; [mag]},$$
$${\rm \BCH= \Mbol(DUSTY)-(H-1.766\times A_{\rm K_{\rm s}}(int)- DM)\; [mag]},$$
$${\rm \BCKs= \Mbol(DUSTY)-( K_S-A_{\rm K_{\rm s}}(int)-DM)\; [mag]},$$
where  DM is the distance modulus, and $J$, $H$, and \Ks\ are the 2MASS magnitudes.
These  empirical BCs are plotted versus the \Teff\ values in Fig.\
\ref{fig_bc}.  For \Teff\ larger than 3300 K and $\tau_{2.2} < 0.08$, 
the  empirical \BCJ\ values appear almost 
constant, with a mean value of 1.73 mag ($\sigma=0.05$ mag) 
and a median value of 1.73 mag. In contrast, 
the theoretical  \BCJ\ has an average of 1.62 mag with a  $\sigma=0.08$ mag.

The  empirical \BCKs\ increases with decreasing temperature, 
 but its slope is steeper than that derived from the theoretical models.
At 3400 K, the \BCKs\ is 0.07 mag larger than the theoretical value, while
at 4100 K, it is 0.12 mag smaller than the theoretical value. 
The stars with visible infrared excess   are marked and deviate 
from the relation for naked stars. Since for naked stars the 
BC$_{\rm J}$ values have 
little dependence on temperatures, the BC$_{\rm J}$ values
clearly decrease with increasing \Aks(env)
 in  Fig. \ref{bctau}.
When estimating the \Mbol\ with the \BCJ,
one can underestimate the \Mbol\ up to 0.6 mag.

\addtocounter{table}{+1}
\begin{table}
\caption{\label{allardBC}  BCs for naked stars were derived with the NextGen models 
([M/H]=0, log(g)=0.5) of \citet{allard11}.}
\begin{tabular}{rrrrr}
\hline
\hline
\Teff & BC$_{\rm J}$ & BC$_{\rm H}$ & BC$_{\rm Ks}$& \\
$\rm [K]$   & [mag]        &[mag]         &[mag]\\
\hline
2600&  2.140&  2.980&  3.221\\
2700&  2.077&  3.002&  3.247\\
2800&  2.013&  2.998&  3.254\\
2900&  1.975&  2.982&  3.248\\
3000&  1.932&  2.960&  3.234\\
3100&  1.890&  2.924&  3.208\\
3200&  1.857&  2.870&  3.152\\
3300&  1.818&  2.813&  3.089\\
3400&  1.762&  2.740&  3.006\\
3500&  1.727&  2.673&  2.924\\
3600&  1.696&  2.608&  2.845\\
3700&  1.668&  2.545&  2.766\\
3800&  1.645&  2.484&  2.690\\
3900&  1.626&  2.426&  2.617\\
4000&  1.611&  2.370&  2.548\\
4100&  1.595&  2.316&  2.480\\
4200&  1.578&  2.262&  2.414\\
4300&  1.558&  2.209&  2.349\\
4400&  1.537&  2.156&  2.284\\
4500&  1.513&  2.103&  2.221\\
\hline
\end{tabular}
\end{table}

\begin{figure}
\begin{center}
\resizebox{0.99\hsize}{!}{\includegraphics[angle=0]{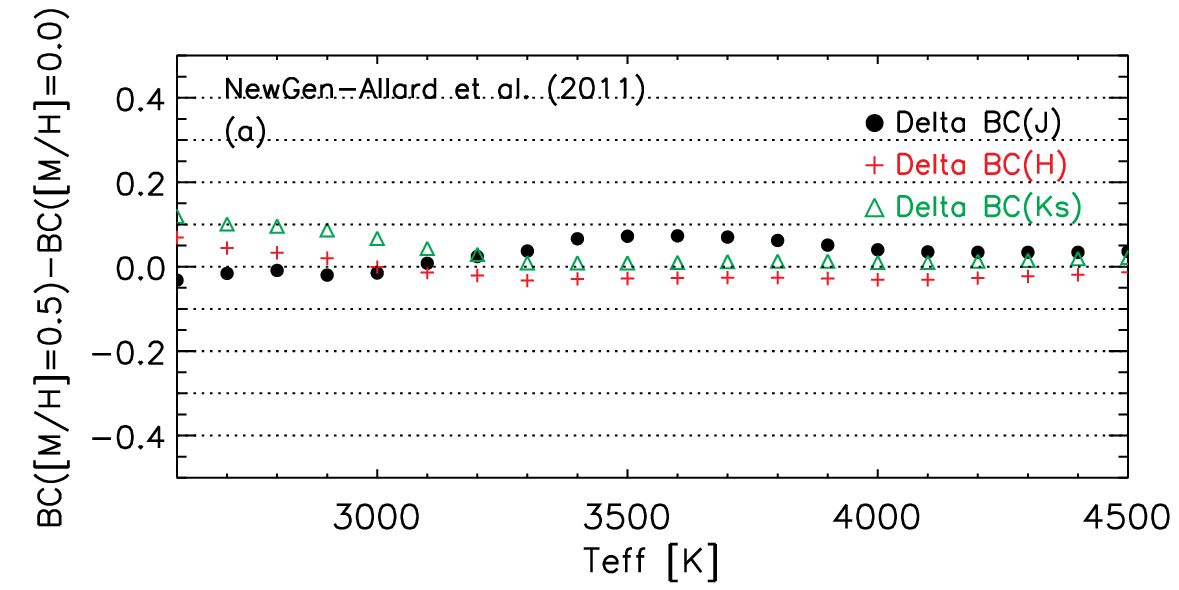}}
\resizebox{0.99\hsize}{!}{\includegraphics[angle=0]{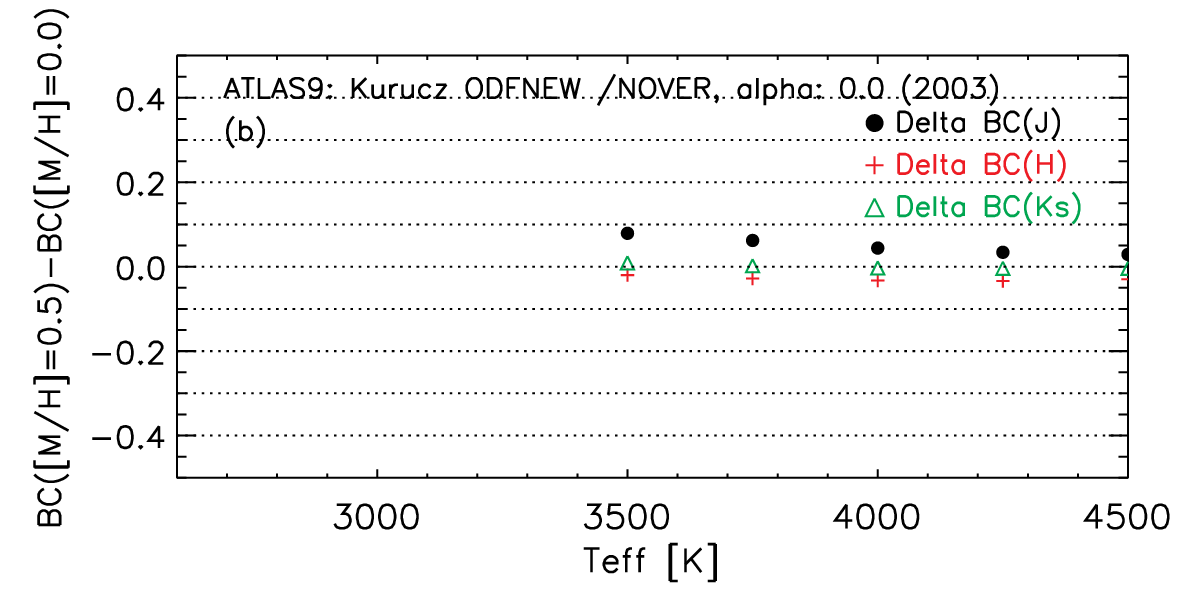}}
\resizebox{0.99\hsize}{!}{\includegraphics[angle=0]{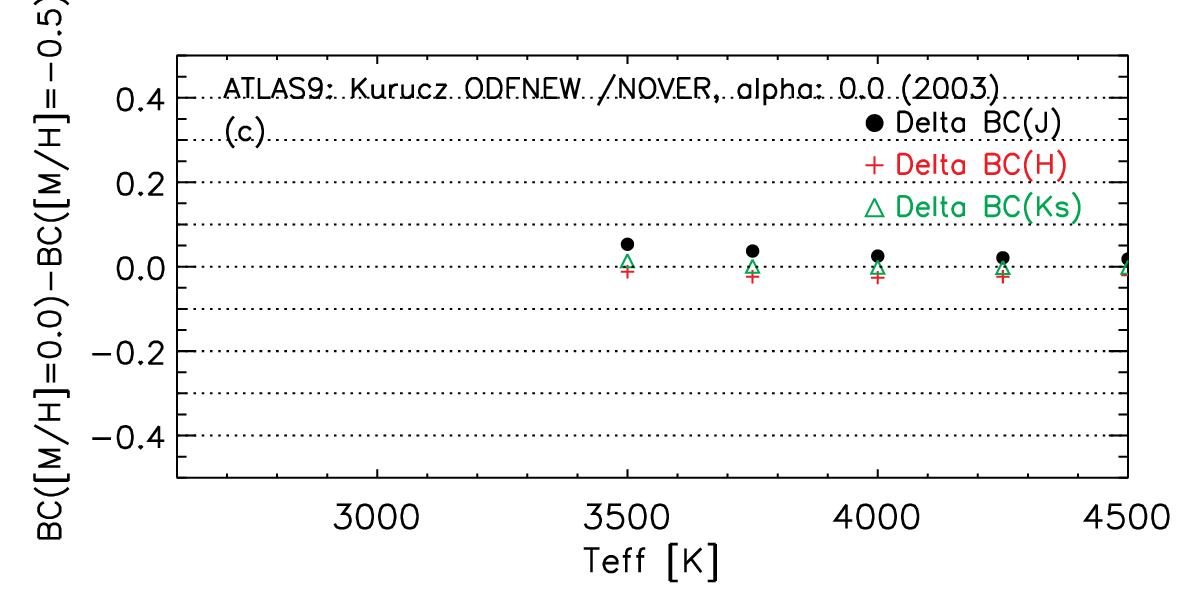}}
\resizebox{0.99\hsize}{!}{\includegraphics[angle=0]{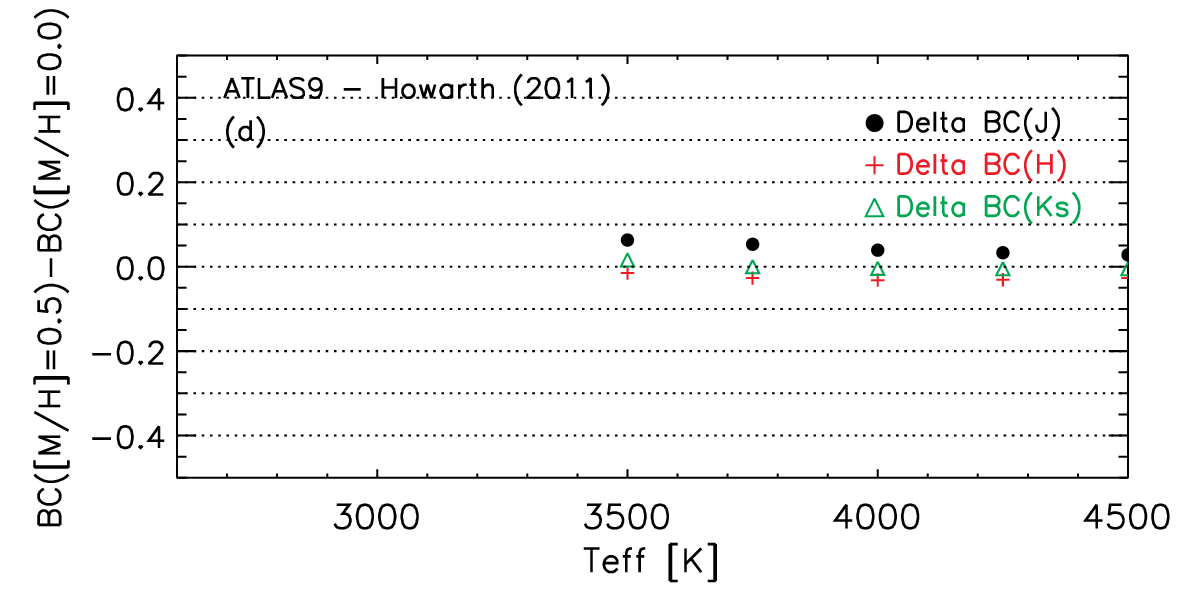}}
\end{center}
\caption{\label{fig.bcmet}
{\it Panel (a):}
Differences between BCs in \( J \), \( H \), and \( K_s \) bands
 inferred from the NextGen models
of \citet{allard11} with 
log$_{10}$(g)=0.5, [M/H]=0.5 dex, and \Teff\ ranging from 2600 to 4500 K
in steps of 100 K; and the BCs inferred from the same set of models, but
with [M/H]=0.0 dex.
{\it Panel (b):}
Differences between BCs  inferred from 
ATLAS9 models with log$_{10}$(g)=0.5, [M/H]=0.5 dex, and 
\Teff=3500, 3750, 4000, 4250, and 4500 K;
and the BCs inferred from the same set of models, but with
[M/H]=0.0 dex.
{\it Panel (c):}
Differences between BCs  inferred from 
ATLAS9 models with log$_{10}$(g)=0.5, [M/H]=0.0 dex, and 
\Teff=3500, 3750, 4000, 4250, 4500 K;
and the BCs inferred from the same set of models, but with
[M/H]=$-0.5$ dex.
{\it Panel (d):} Similar to Panel (b); this time,  the 
ATLAS9 models are those distributed by \citet{howarth11}.
Dotted horizontal lines are drawn every 0.1 mag for easier visualization. 
}
\end{figure}

The above description is based on  \Mbol\ values from synthetic spectra 
with solar metallicity.
The theoretical dependence of the  BC$_{\rm J}$, 
BC$_{\rm H}$, and BC$_{\rm Ks}$ on the metallicity is also explored. 
In Fig.\ \ref{fig.bcmet}, the differences between BCs from models
at [M/H]=+0.5 and those from models at [M/H]=+0.0 are displayed. 
For the NextGen models (panel a),
$\Delta$  BC$_{\rm J}$ =  0.05 mag with  
$\sigma= 0.02$ mag when considering \Teff\ from 3200 to 4500 K;
$\Delta$  BC$_{\rm H}$ =  $-$0.03 mag with  
$\sigma= 0.01$; and 
$\Delta$  BC$_{\rm Ks}$ =  0.014 mag with  
$\sigma= 0.01$. 
For the ATLAS9 models (panel b and d) 
the variations are:
$\Delta$  BC$_{\rm J}$ =  0.050 mag with  
$\sigma= 0.021$ mag when considering \Teff\ from 3200 to 4500 K;
$\Delta$  BC$_{\rm H}$ =  $-$0.029 mag with  
$\sigma= 0.006$ mag; and 
$\Delta$  BC$_{\rm Ks}$ =  $-$0.000 mag with  
$\sigma= 0.006$ mag. 
For the ATLAS9 models, similar differences are calculated
when comparing models at [M/H]=+0.0 dex and  
models at [M/H]=$-0.5$ dex 
(which are not available in Allard’s NextGen grid).

In summary,  uncertainties in BCs  due to metallicity can be 
as high as 0.1 mag. Additionally, 
uncertainties in BC$_{\rm Ks}$  resulting from uncertain temperatures 
are of a similar magnitude 
(e.g., 0.12+0.08 mag for \Teff=4100 K).

\subsection{Isolation versus clustering}
\label{sec.cluster}
In the present work, we individually selected cRSGs on the basis
of their color and luminosities. 
Due to their young age (5-40 Myr), RSGs are found 
to be associated with giant molecular  complexes 
\citep[e.g.,][]{messineo14a,messineo21}; however,
at their ages, most of the stellar clusters 
have probably already been dissolved. 
They must therefore be detected individually, and the Gaia
parallax allows us to do this.

Most of the efforts of the past two decades were focused
on identifying  clusters rich in RSGs. 
Until 2006, the only clusters known with four and five RSGs
were Westerlund 1 and NGC 7419 \citep[e.g.,][]{clark05, caron03}.
From 2006 to 2012, marvelous and rare stellar clusters rich in RSGs 
(RSGC1, RSGC2, RSGC3, Alicante7, Alicante8)
were detected in the Milky Way
between longitudes 25$^\circ$ and 35$^\circ$ at a 
heliocentric distance of about 6 kpc,
located at the near end side of the Galactic bar
\citep{figer06,davies07,clark09b,messineo14b,chun24}.
The strict selection of parallax quality used in the 
present work excluded any member of 
these three distant clusters rich in RSGs located at 
the near-end side of the Galactic bar.

Eighteen percent of the newly selected cRSGs are found to be cluster members. 
Each stellar position was searched in VIZIER, and the appearing catalogs 
of cluster members were annotated.
Then, the memberships were re-retrieved automatically 
by combining the matches found in the following catalogs:
\citet{cantat20},
\citet{castro22},
\citet{tar22},
\citet{cantat18},
\citet{vanGroeningen23},
\citet{hunt24},
\citet{kos24},
\citet{chiIII23},
\citet{qin23},
\citet{he23},
\citet{heb23},
\citet{he22a},
\citet{he22b},
\citet{hao22},
\citet{ore21},
\citet{dias18}, and
\citet{samp17}.
If matches were found in multiple catalogs, only the first was retained
(following the above listing) and only if the probability 
of membership was higher than 60\%.
All used catalogs, except for \citet{samp17}, are based on Gaia data.
The cluster membership is annotated in Table 1.

Theoretically, the precise age of a simple stellar population 
hosting RSGs depends on the inclusion and strength of stellar rotation.
For non-rotating models, RSGs 
are found to populate clusters with ages from 4.5 to 40 Myr
\citep[e.g.,][]{messineo11}.
Observationally, small variations in age may arise from 
the selected sample of members and average parallax.
Cluster parameters were taken from
the compilation  by \citet{perren23}, 
as well as from the VIZIER database.
For each cluster, the two most recent determinations of age and distance
were collected. 
For 58\% of  cRSGs associated with a cluster, the  cluster  ages are consistent 
with the hosting of an RSG; i.e., they are younger than 41 Myr.
For 40\%, the associated clusters are older than 41 Myr; 
all but three cRSGs are younger than 300 Myr.
The selection seems to yield $\approx 60$\% of authentic RSGs.
When excluding members from the work of \citet{samp17}, which is
not based on Gaia parallaxes, this percentage rises to 66\%.

A full investigation of the stellar clusters is beyond the scope of this paper.
The names of the clusters and their ages are listed in the target list (Table 1).

\begin{figure*}
\begin{center}
\resizebox{0.49\hsize}{!}{\includegraphics[angle=0]{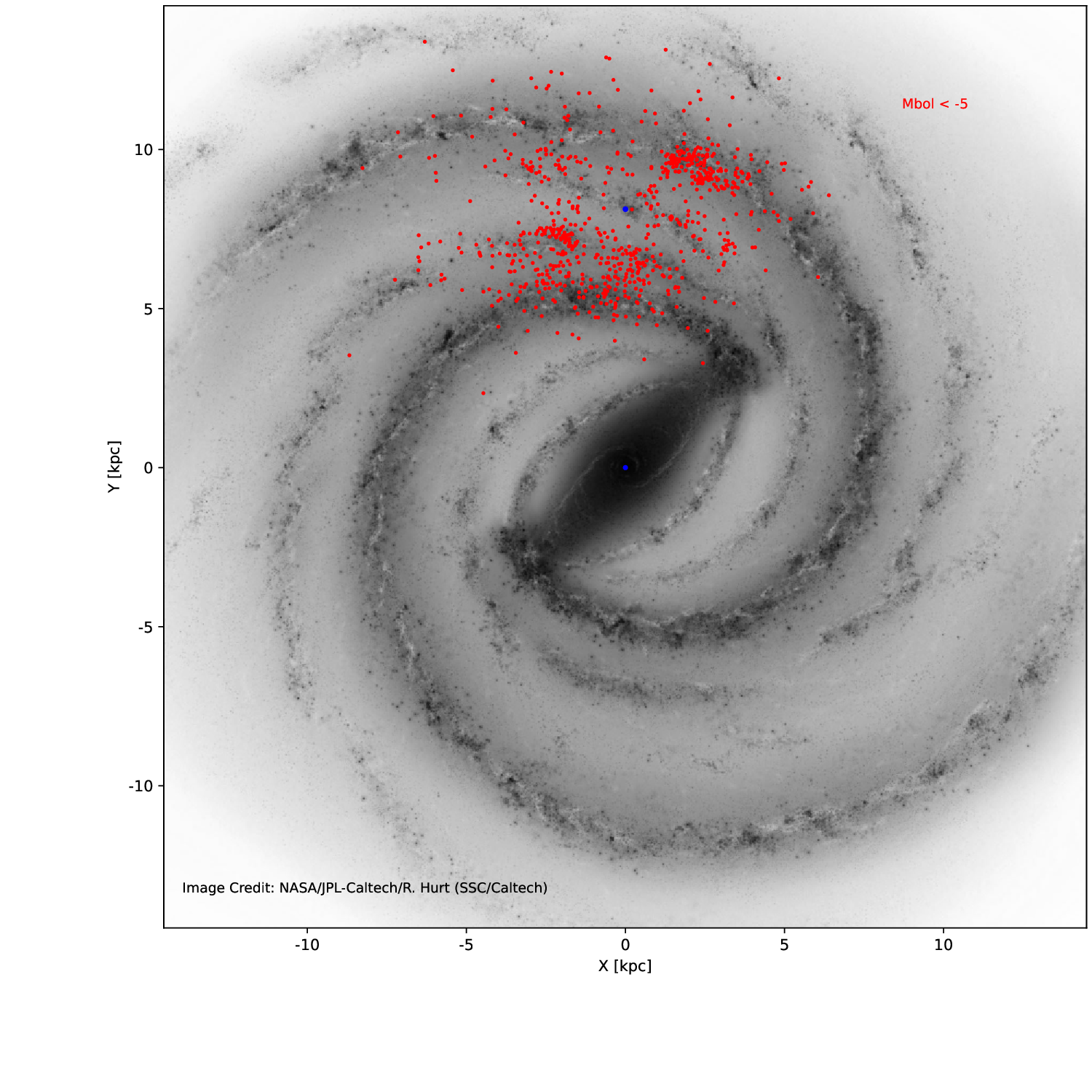}}
\resizebox{0.49\hsize}{!}{\includegraphics[angle=0]{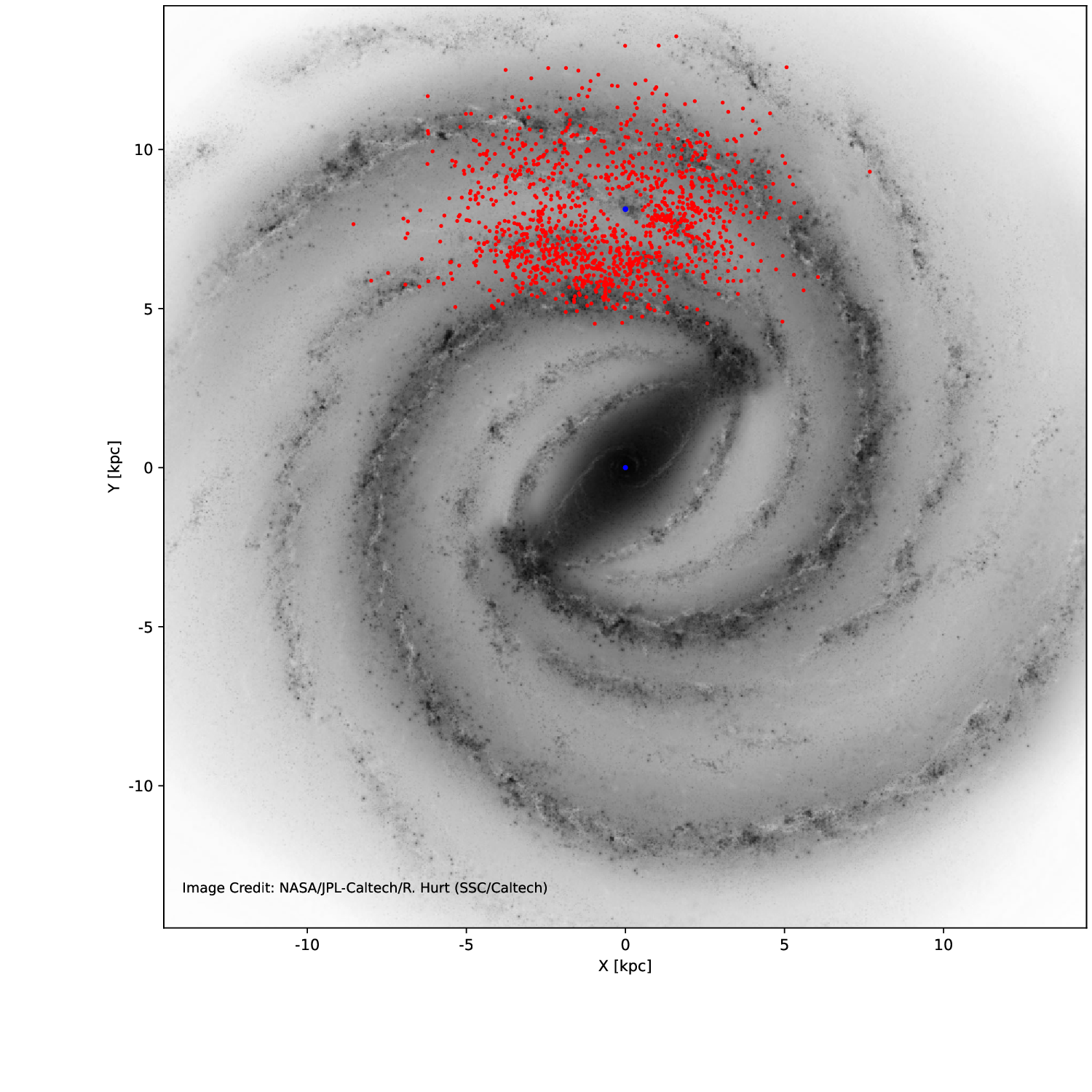}}
\end{center}
\caption{
\label{xydistributionhurt}
{\it Right panel:} XY view of  stars  located in areas A and B 
 of \Mbol\ versus \Teff\ diagram of  \citet{messineo19} 
and newly selected Gaia-2MASS cRSGs, which are also located in areas A and B 
(red dots).  The background grayscale image is the artistic XY 
view of the Galactic plane illustrated by Dr. Hurt.
{\it Left panel:} 
Red data points show the complementary dataset  
(2122 stars with equal brightness
 (\Mk $< -8.5$ mag), which were excluded from the cRSGs).
}  
\end{figure*}

\begin{figure}
\begin{center}
\resizebox{0.99\hsize}{!}{\includegraphics[angle=0]{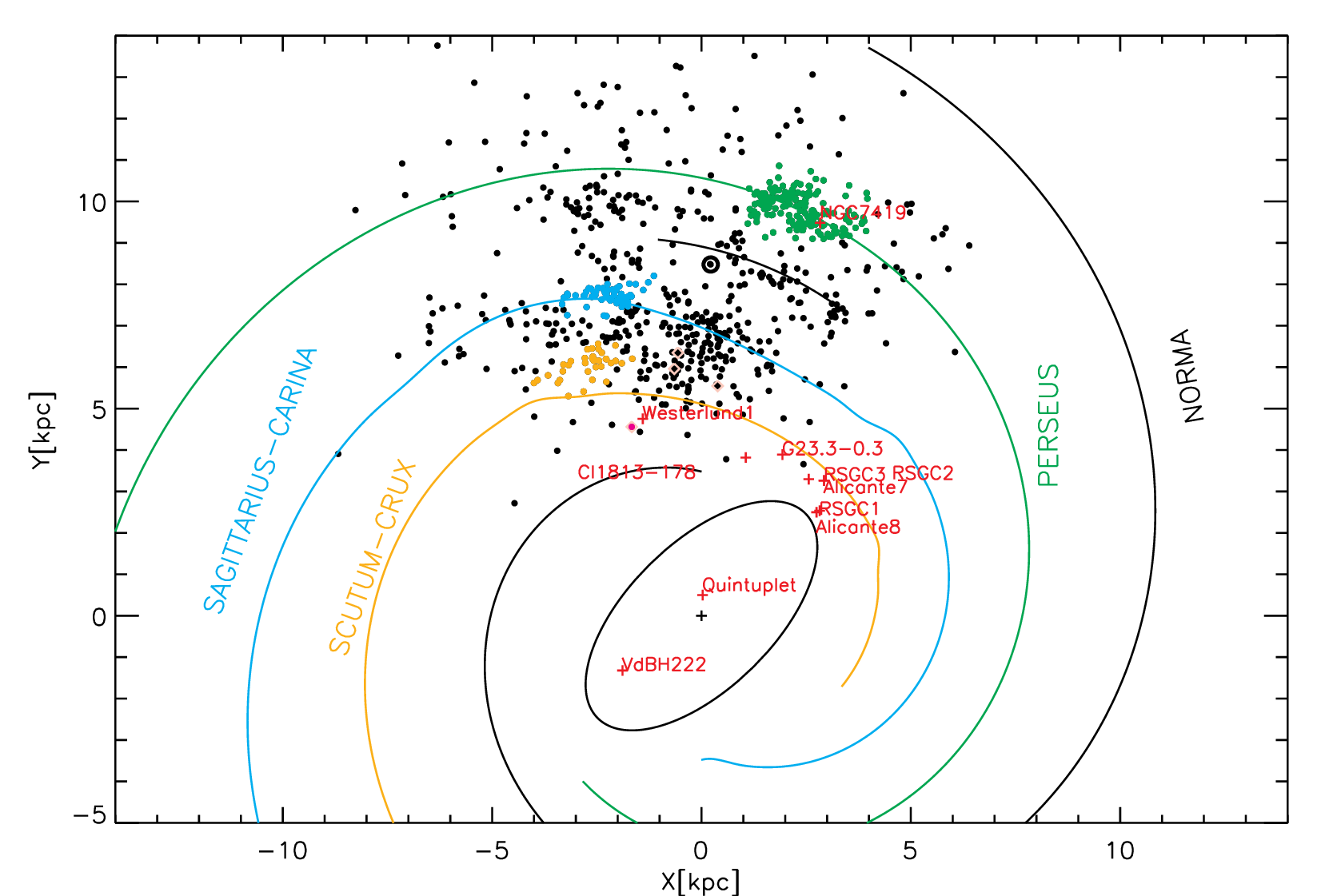}}
\resizebox{0.99\hsize}{!}{\includegraphics[angle=0]{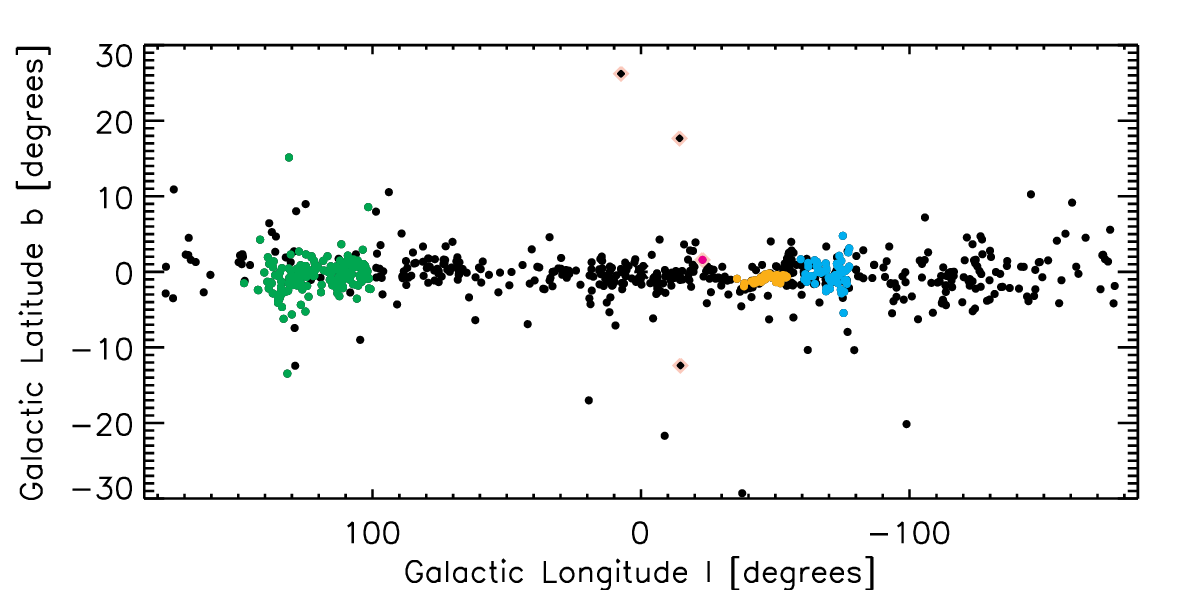}}
\resizebox{0.99\hsize}{!}{\includegraphics[angle=0]{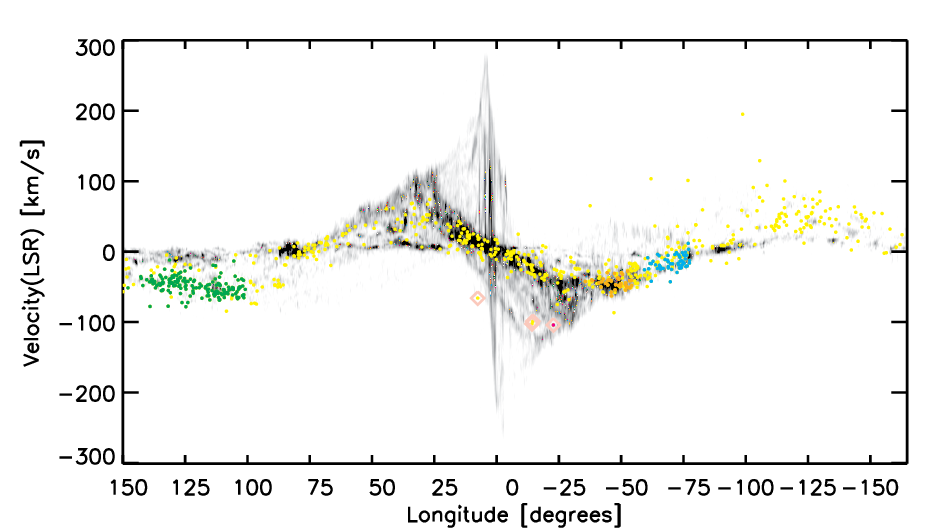}}
\end{center}
\caption{\label{xydistribution} \label{xydistributionGMC}
{\it Upper panel:} XY view of cRSGs from areas A and B of \citet{messineo19},
along with new Gaia-2MASS cRSGs.
 A distance of 8.5 kpc is adopted, as in the work of \citet{ocker24}.
Green filled circles indicate  stars located on  the large over-density of RSGs  of 
the Perseus   arm.
Cyan filled circles mark  stars   corresponding to 
an apparent concentration  of RSGs 
 at the tangent point of the Sagittarius-Carina 
arm  at l$\approx -78^\circ$ \citep[][]{hou14}.
Orange-filled circles indicate stars located  at the tangent point
of the Scutum-Centaurus arm
 \citep[at l$\approx -50^\circ$,][]{hou14}.
Known massive clusters ($>10^4$ \Msun) rich in RSGs are marked in red.
 The  spiral arms of \citet{cordes02},  
taken from the native NE2001p Python code of \citet{ocker24}, are also shown.
{\it Middle panel:} Longitudes versus latitudes  of cRSGs.
The green, cyan, and orange data points are as specified in the upper panel.
{\it Lower panel:} 
 \vlsr\ velocities versus longitudes  of cRSGs  (shown in yellow).
The green, cyan, and orange data points are as specified in the upper panel.
The background grayscale image is the CO map by \citet{dame01}.
Four points with  peculiar velocities are indicated with pink diamonds;
three of them are at high latitudes. 
}  
\end{figure}

\section{XY distribution  and longitude-velocity diagram}
\label{sec.distribution}

Asymptotic giant branch stars, being old, are good tracers of the Galactic potential,
and, for example, their kinematics have been studied to characterize the 
central Galactic bar  \citep[e.g.,][]{habing06,tian24}.
On the contrary, RSGs are young and their distribution and kinematics
give a picture of the Galactic locations where  
massive star formation occurred between 5 and 30-40 Myr ago; 
there is approximately one RSG for every 
burst of 10,000 stars \citep[e.g.,][]{clark09}.
That is, RSGs are young enough to still mark sites of violent
cloud collisions and gas compression, and they populate
the spiral arms. Galaxies are 
far from precise mathematical logarithmic spirals.
Spiral arms are transitional phenomena, forms, and reforms. 
A closer look at the pictures of grand design spirals shows a 
richness of arms and connecting ridges.
In the Milky Way, discussion is ongoing with regard to the actual 
number of spiral arms; it may be from two to four \citep[][]{hou14}, 
or perhaps there are two that bifurcate.

In Fig. \ref{xydistributionhurt}, the best sample of cRSGs, 
individually selected from Gaia, is overlaid on an 
artistic representation of the Galactic plane based on Spitzer data by Dr. 
Hurt\footnote{
\href{https://science.nasa.gov/resource/the-milky-way-galaxy/}{the-milky-way-galaxy}}; 
yet, in  Fig. \ref{xydistribution}
the sample of highly probable RSGs 
is plotted in the XY plane, in the longitude
and latitude diagram, and in the velocity-versus-longitude diagram. 
We  considered a sample  of 762 bright late-type stars
that are located in areas A and B of the luminosity-versus-\Teff\ diagram and with radial velocity.
466 stars are from the literature collection by 
\citet{messineo19}\footnote{For homogeneity, the luminosities and 
areas of stars in \citet{messineo19} 
and \citet{messineo23} were re-estimated with the same extinction 
law adopted here; i.e., using an infrared extinction law with 
index = $-$2.1. They had previously been calculated with 
an index= $-1.9$.},
20 are from \citet{messineo23}, and 276 are from the present work.
For 96\%  of the sample, radial velocities were available in the main 
Gaia DR3 catalog \citep{katz23}. The remaining 4\% of the measurements 
were found in  the compilation of radial velocities 
by \citet{tsantaki22}; this included a few measurements from APOGEE and 
Gaia DR2 velocities (not included in the Gaia DR3 release).
The Gaia barycentric radial velocities were transformed in the 
local standard of rest (LSR) system using
the solar  motion by \citet[model A5,][]{reid19}.
In the XY plane of Fig. \ref{xydistribution}, 
the model of spiral arms made with distance measures of pulsars 
\citep{cordes02} is superimposed.

The new selection adds stars at greater distances from us and deeper into the
inner Galaxy to touch the Scutum-Centaurus arm.
The distribution of cRSGs appears far from being homogeneous;
indeed, there are several over-densities of cRSGs  along the spiral arms,
and filaments of   cRSGs decorate the inter-arm regions. 
In Fig. \ref{xydistribution}, the large over-density on the Perseus spiral arm 
is colored in green ($100^\circ< l <150^\circ$, $1 < Y< 4$ kpc).
 Another concentration of stars,  colored  cyan,
 appears on the Sagittarius arm
at  $-55^\circ > l >-80^\circ$ and  $ 7.3 \la Y< 8.0$ kpc;
in the longitude-velocity diagram, this  concentration appears to be
stretched along the terminal velocity curve, and to end at the 
tangent point\footnote{ ``One can identify some of the
dense emission ridges in the CO longitude-velocity diagram 
with Galactic spiral arms; where these meet the terminal velocity curve, 
they can be recognized as bumps where $\delta$vt/$\delta$l=0...
The inferred spiral arm tangents coincide with features
along the terminal curve'' \citep{englmaier99}.}
\citep[as indicated in Table 2 of][]{hou14}
of the Sagittarius arm  at $l\approx-78^\circ$,  
consistently with the arm modeling of \citet{hou14}.
A narrow feature, colored  orange,  is visible on 
the longitudes-versus-latitudes plot, showing
decreasing latitudes  with 
increasing longitudes 
(between $-35^\circ > l >-55^\circ$ and $-2^\circ < b <0^\circ$). 
In the XY plane, this feature appears as a group of stars
closer to the inner Scutum-Centaurus arm ($Y < 6.7$ kpc). 
In the longitude-velocity diagram, 
these stars lie along the terminal velocity curve and
culminate at the  tangent point of the  
Scutum-Centaurus arm at $l\approx-50^\circ$ \citep[e.g.,][]{englmaier99,hou14}. A filament of cRSGs appears aligned with the Local Arm, another 
filament extends from 
$-5< X<0$ kpc at Y$\approx$ 9 kpc. \\
Despite the small numbers, this appearance
shows a remarkable resemblance with the distribution of giant molecular clouds (GMC)
shown in the work of \citet{hou14}.
The cRSGs are not evenly distributed along the spiral arms,
and they also populate the inter-arm regions.
The two  over-densities centered on the Perseus arm at $l\approx130^\circ$, 
and on the Sagittarius arm at $l\approx-70^\circ$  
appear symmetrically disposed around the Sun   along directions 
where over-densities of GMCs also lie \citep{hou14}.

\section{Summary}
\label{mysummary}

An exercise to select bright late-type stars, cRSGs, from GAIA and 2MASS catalogs
was carried out.
While \citet{messineo19} made a catalog of stars previously classified as 
class I cool stars, \citet{messineo23} looked at the Gaia DR3 
parameters  to locate about 20 additional cRSGs and 
to find out that, for cool bright stars, the extinction and temperatures 
are still highly uncertain.
The photometric diagnostics presented in \citet{messineo23} were 
adopted here to select 335 additional cRSGs, with 282 located in areas A \& B
(39 in area A and 243 in area B). 
Only 12 of these stars are included in the catalog of \citet{healy24}.
Only six of the 335 stars have Gaia DR3 RVS data; however, they belong to areas
E and C. There are no Gaia DR3 RVS data for stars in areas A and B, 
which is our conservative definition of cRSGs.

 A small percentage (18\%) of stars are found to be members 
of  stellar clusters. While 58\% of the clusters are reported 
to be younger than 41 Myr old, most have ages below 300 Myr.
By extrapolating these percentages to the entire sample,
58\% should be true RSGs.

In the XY plane, the cRSGs appear to follow the spiral arms
and to populate inter-arm regions. The 
known Perseus over-density of RSGs is  clearly visible. 
The velocity-longitude diagram suggests that two other over-densities of the XY plane
are related to tangent points.
A number of cRSGs appear to be aligned with the Local Arm.

\section{Data availability}
Table 1, which lists  the 335 new selected late-type stars, 
is only available in electronic form at the CDS 
via anonymous ftp to cdsarc.u-strasbg.fr (130.79.128.5) or 
via http://cdsweb.u-strasbg.fr/cgi-bin/qcat?J/A+A/
(check  also at  \href{https://mariamessineo.github.io/rsg_Ak/}{mariamessineo.github.io/rsg\_Ak}).

\begin{acknowledgement}
This work has made use of data from the European Space Agency (ESA) mission {\it Gaia}
($http://www.cosmos.esa.int/gaia$), processed by the {\it Gaia} Data Processing and Analysis
Consortium (DPAC, $http://www.cosmos.esa.int/web/gaia/dpac/consortium$). Funding for the DPAC
has been provided by national institutions, in particular the institutions participating in the {\it
Gaia} Multilateral Agreement. 
This publication makes use of data products from the Two Micron All
Sky Survey, which is a joint project of the University of Massachusetts and the Infrared Processing
and Analysis Center / California Institute of Technology, funded by the National Aeronautics and Space
Administration and the National Science Foundation. 
This work is based on observations made with
the Spitzer Space Telescope, which is operated by the Jet Propulsion Laboratory, California
Institute of Technology under a contract with NASA.
This research made use of data products from the Midcourse Space Experiment, the processing of which
was funded by the Ballistic Missile Defense Organization with additional support from the NASA
office of Space Science. 
This publication makes use of data products from WISE, which is a joint
project of the University of California, Los Angeles, and the Jet Propulsion Laboratory / California
Institute of Technology, funded by the National Aeronautics and Space Administration. 
This research has made use of the VizieR catalogue access tool, 
CDS, Strasbourg, France, and SIMBAD database.
This research utilized  the NASA’s Astrophysics Data System Bibliographic Services.\\ 
We thank the Python, Astropy, NumPy, Pandas, Pivo, SciPy, Matplolib 
communities for providing their packages.  
We thank Dr. Hurt for his illustrations of the Milky Way.
This work uses the RSG catalog by Messineo M. \& Brown A. (2019) which was
supported by the National Natural Science Foundation of China
(NSFC-11773025, 11421303), and USTC grant KY2030000054. 
MM thanks Prof. Anthony Brown for helping her to realize the 2019 catalog,
and  Dr. Schuyler D. Van Dyk  for appreciating her infrared catalog; this fact 
motivated this  additional search for cRSGs. Another acknowledgment goes 
to Prof. Frank Bertoldi for discussion on interstellar extinction.
\end{acknowledgement}

\end{document}